\documentclass{jfm}
\usepackage{epsfig}
\usepackage{amsmath}
\usepackage{natbib}
\usepackage{bm}
\usepackage{epsf}

\usepackage{graphicx}
\usepackage{dcolumn}
\usepackage{bm}
\usepackage{edit}
\usepackage{localdefs}

\newcommand{\Ca}{{\rm Ca}}
\def\etal{\mbox{\it et al.\ }}

\begin{document}

\title{Hysteretic and chaotic dynamics of viscous drops in creeping
flows with rotation}
\shorttitle{Hysteretic and chaotic dynamics of viscous drops}

\author[Young, B{\l}awzdziewicz $\etal$] {Y.-N. Young$^1$, Jerzy
B{\l}awzdziewicz$^2$, V. Cristini$^3$ and R. H. Goodman$^1$}
\affiliation{ $^1$Department of Mathematical Sciences, New Jersey
Institute of Technology, Newark, New Jersey, 07102-1982, USA\\ $^2$Department of Mechanical
Engineering, Yale University, P.O. Box 20-8286, New Haven, Connecticut,
06520-8286, USA\\ $^3$School of Health Information, University of
Texas, Houston, USA}

\maketitle
\begin{abstract}

It has been shown in our previous publication
\citep{Blawzdziewicz-Cristini-Loewenberg:2003} that high-viscosity
drops in two dimensional linear creeping flows with a nonzero
vorticity component may have two stable stationary states. One state
corresponds to a nearly spherical, compact drop stabilized primarily
by rotation, and the other to an elongated drop stabilized primarily
by capillary forces. Here we explore consequences of the drop
bistability for the dynamics of highly viscous drops. Using both
boundary-integral simulations and small-deformation theory we show
that a quasi-static change of the flow vorticity gives rise to a
hysteretic response of the drop shape, with rapid changes between the
compact and elongated solutions at critical values of the
vorticity. In flows with sinusoidal temporal variation of the
vorticity we find chaotic drop dynamics in response to the periodic
forcing.  A cascade of period-doubling bifurcations is found to be
directly responsible for the transition to chaos.  In random flows we
obtain a bimodal drop-length distribution. Some analogies with the
dynamics of macromolecules and vesicles are pointed out.
\end{abstract}

\section{Introduction}

Investigations of dynamical properties of fluid--fluid dispersions
(e.g., emulsions \citep{Borwankar-Case:1997,Mason:1999} and polymer
blends 
\citep{%
Tucker_III-Moldenaers:2002,%
Windhab-Dressler-Feigl-Fischer-Megias_Alguacil:2005%
}) 
require detailed understanding of the behavior of viscous drops in
creeping flows.  Such understanding is also crucial in development of
new drop-based microfluidic systems 
\citep{%
Whitesides-Stroock:2001,%
Tan-Fisher-Lee-Cristini-Lee:2004,%
Song2006,%
Grigoriev2006%
}.
Therefore, drop dynamics at small Reynolds
numbers have been extensively studied experimentally
\citep*{%
Torza-Cox-Mason:1972,%
Bentley-Leal:1986,%
Bigio-Marks-Calabrese:1998a,%
Guido-Minale-Maffettone:2000,%
Cristini-Guido-Alfani-Blawzdziewicz-Loewenberg:2003,%
Guido-Grosso-Maffettone:2004%
},
computationally 
\citep{%
Rallison-Acrivos:1978,%
Kennedy-Pozrikidis-Skalak:1994,%
Cristini-Blawzdziewicz-Loewenberg:1998a,%
Zinchenko-Rother-Davis:1999,%
Cristini-Blawzdziewicz-Loewenberg:2001,%
Cristini-Guido-Alfani-Blawzdziewicz-Loewenberg:2003,%
Renardy:2006%
}
and theoretically
\citep{%
BarthesBiesel-Acrivos:1973,%
Rallison:1980,%
Blawzdziewicz-Cristini-Loewenberg:2002a,%
Blawzdziewicz-Cristini-Loewenberg:2003,%
Vlahovska-Loewenberg-Blawzdziewicz:2005%
}.

These investigations revealed complex nonlinear drop dynamics
resulting from the coupling between the drop shape and fluid flow.
Examples of nonlinear phenomena that occur under
creeping-flow conditions include formation of self-similar neck
regions during a drop breakup process
\citep{%
Blawzdziewicz-Cristini-Loewenberg:1997,%
Lister-Stone:1998%
},
universal slow evolution of drops near the critical flow strength
above which there are no stationary drop shapes
\citep{%
Blawzdziewicz-Cristini-Loewenberg:1998,%
Navot:1999,%
Blawzdziewicz-Cristini-Loewenberg:2002a%
},
and existence of two branches of stable stationary shapes of highly
viscous drops in two-dimensional Stokes flows with nonzero vorticity
\citep{Blawzdziewicz-Cristini-Loewenberg:2003}.

As revealed by the analysis presented by
\cite{Blawzdziewicz-Cristini-Loewenberg:2003}, there exists a
flow-parameter range where a high-viscosity drop can either adopt a
nearly spherical shape stabilized primarily by the rotational flow
component or an elongated shape stabilized primarily by capillary
forces.  Abrupt changes of the drop shape from one state to the other
can be used in manipulation of emulsion micro-structure and for
controlling the behavior of highly viscous drops in microfluidic
devices.  Due to discontinuous changes of emulsion microstructure, the
bistable drop behavior may also significantly affect emulsion
rheology.  Moreover, the mechanism of the bistability is of
fundamental interest because of close analogies to the dynamics of
vesicles 
\citep{Misbah:2006,%
Mader-Vitkova-Abkarian-Viallat-Podgorski:2006,%
Vlahovska-Gracia:2007} 
and macromolecules \citep{Blawzdziewicz:2006} in
external flows.

Due to the fundamental significance and because of potential
applications (such as those mentioned above) it is important to
explore the dynamics of highly viscous drops in linear flows with
nonzero vorticity.  However, the investigations presented by
\cite{Blawzdziewicz-Cristini-Loewenberg:2003} were limited to
stationary drop shapes and stationary external flows.  In the present
study we focus on drop behavior in time-dependent flows.  We elucidate
the physical mechanism that give rise to the bistable drop behavior
and examine the consequence of these mechanism for drop response to
time variation of the fluid vorticity.

The system dynamics is investigated via direct boundary-integral
simulations 
\citep{Pozrikidis:1992,%
Cristini-Blawzdziewicz-Loewenberg:2001,%
Blawzdziewicz:2006} 
and by using a small-deformation approach
\cite*{Vlahovska:2003,%
Vlahovska-Loewenberg-Blawzdziewicz:2005}.  
In particular we
show that the small-deformation equations with only several essential
terms retained reproduce complex dynamical features of drop evolution
that are associated with drop bistability.

To emphasize important aspects of the drop dynamics we consider three
flow variation protocols.  In the first protocol, the vorticity is
slowly increased and then decreased.  We find that such quasistatic
vorticity ramping gives access to both the elongated and compact,
nearly spherical stationary drop shapes.  The drop exhibits a
hysteretic behavior, with transitions between the compact shape
(rotationally stabilized) and elongated shape (stabilized by capillary
forces) occurring at different values of the vorticity when it is
slowly ramped up or down.

In the second protocol, the vorticity undergoes finite-frequency
periodic oscillations.  As expected, at low frequencies the drop
behavior is quasistatic, with a hysteresis loop analogous to the one
observed for linear ramping.  At high frequencies the oscillations
average out, and the drop undergoes small oscillations around the
stationary shape corresponding to the average flow.  However, at
intermediate frequencies we find a much more complex behavior.  In
particular we show that there exists a frequency and amplitude domain
where the drop response to the periodic forcing is chaotic.  Since in
the creeping flow regime fluid motion is governed by the linear Stokes
equations, the nonlinear chaotic behavior of the drop stems entirely
from the coupling of the drop shape to the fluid velocity.  An
analysis of drop motion in the chaotic domain indicates that the
transition to chaos is associated with the existence of two stationary
states observed in steady flow.

In our third flow-variation protocol, the vorticity of the imposed
flow undergoes random changes.  We observe, that the resulting
statistical distribution of the drop length is bimodal in a certain
regime of flow parameters, with two peaks around the drop length
corresponding to the short and long stationary solutions.  Hence, we
find that also in this problem the existence of two stationary states
underpins drop behavior in a time-dependent flow.

In \S\,\ref{Viscous drop in linear flows} the system considered in our
paper is defined.  The quasistatic hysteretic drop behavior is
analyzed in \S\,\ref{Hysteretic behavior}, our results for chaotic
drop dynamics are presented in \S\,\ref{sec:sinusoidal_chaos}, and
drop motion in linear flows with randomly varying vorticity is
discussed in \S\,\ref{Drop Statistics in a linear flow with Stochastic
Vorticity}.  Our findings are summarized in \S\,\ref{sec:conclusion}.

\section{Viscous drops in two-dimensional linear flows}
\label{Viscous drop in linear flows}

We consider a viscous drop suspended in an incompressible Newtonian
fluid of a constant viscosity $\mu$.  The viscosity of the drop fluid
is $\hat\mu=\lambda\mu$, and the interfacial tension between the two
phases is $\sigma$.  The drop is surfactant free, and no Marangoni
stresses are present.  There are also no buoyancy forces.  We focus
here on nonlinear effects that stem entirely from the coupling of the
fluid flow to the drop shape (but not from the fluid
inertia). Therefore, the creeping-flow conditions are assumed.

In the creeping-flow regime the fluid motion in the regions inside
($\mu_i=\hat\mu$) and outside ($\mu_i=\mu$) the drop is governed by
the Stokes equations
\begin{eqnarray}
\label{eq:stokes01}
\mu_i\nabla^2\bu&=&\boldsymbol{\nabla} p,\\
\label{eq:incomp}
\nabla \cdot\bu &=&0.
\end{eqnarray}
The fluid velocity $\bu$ is continuous at the drop interface
$\Omega$.  Due to the absence of the Marangoni stresses the tangential
viscous traction is also continuous.  The jump in the normal viscous
traction  across $\Omega$ is equal to the capillary pressure
\begin{equation}
\label{eq:stressbalance}
[\unitNormal\cdot\boldsymbol{\tau}\cdot\unitNormal] = 2\kappa\sigma,
\end{equation}
where $\boldsymbol{\tau}$ is the viscous stress tensor, 
$\unitNormal$ is the outward normal unit
vector, and $\kappa$ is the local curvature.  

\begin{figure}
\begin{center}{\includegraphics[width=4.0in]{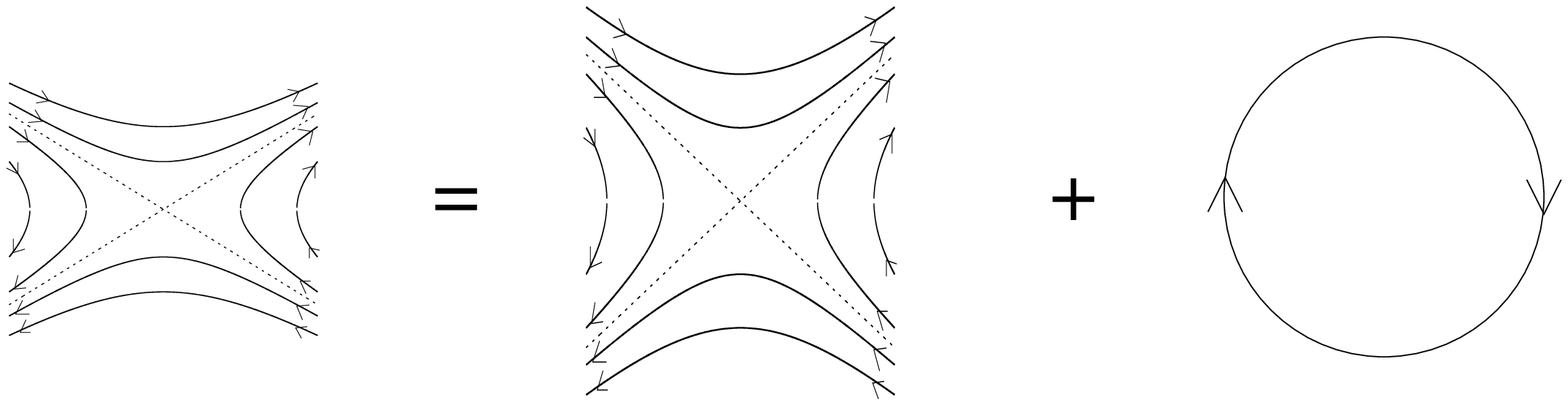}}\end{center}
\begin{picture}(300,8)(0,0)
\put(75,0){$\bu_0$}
\put(176,0){$\dot\gamma\,\strainTensor\bcdot\br$}
\put(290,0){$\dot\gamma\beta\,\rotationTensor\bcdot\br$}
\end{picture}
\caption{Decomposition of a linear incident flow into pure 
strain and rigid-body rotation. 
}
\label{flow decomposition}
\end{figure}

The drop is subject to two-dimensional linear incident flow
\begin{equation}
\label{decomposition of external flow}
\bu_0(\br)=\dot\gamma(\strainTensor+\beta\rotationTensor)\bcdot\br,
\end{equation}
where $\dot\gamma$ is the strain rate, $\beta$ is the dimensionless
vorticity parameter, $\br$ is the position, 
and $\strainTensor$ and $\rotationTensor$ are the
symmetric and antisymmetric parts of the normalized velocity-gradient
tensor.  In an appropriately adopted coordinate system we have
\begin{equation}
\label{strain and vorticity tensors}
\strainTensor=\frac{1}{2}\left(\begin{array}{ccc}
                             0 & 1 & 0 \\
			     1 & 0 & 0 \\
			     0 & 0 & 0
			     \end{array}\right),
\qquad
\rotationTensor=\frac{1}{2}\left(\begin{array}{ccc}
                             0 & 1 & 0 \\
			     -1 & 0 & 0 \\
			     0 & 0 & 0
			     \end{array}\right),
\end{equation}
 without a loss of generality.  Accordingly, $\beta=0$ corresponds to
a purely straining flow with the extensional axis $x=y$ and the
compressional axis $x=-y$, and $\beta=1$ corresponds to shear flow in
the $x$ direction with the velocity gradient in the $y$ direction.
The tensor $\rotationTensor$ in equation \refeq{decomposition of
external flow} describes the rigid-body rotation in the anti-clockwise
direction with the angular velocity
\begin{equation}
\label{angular velocity}
\omega=\halff\beta\dot\gamma.
\end{equation}
The decomposition of the flow field \refeq{decomposition of external flow}
into the straining and rotational components
\refeq{strain and vorticity tensors} is sketched in figure \ref{flow
decomposition}.

The dynamics of our system is characterized by three dimensionless
parameters: There is the viscosity ratio $\lambda$ that characterizes
dissipative forces in the drop- and continuous-phase fluids.  The
capillary number
\begin{equation}
\label{capillary number}
\Ca=\frac{a\mu\dot\gamma}{\sigma}
\end{equation}
(where $a$ is the radius of an undeformed drop) characterizes the
ratio between the deforming viscous forces produced by the imposed
flow \refeq{decomposition of external flow} and the capillary forces
that resist drop deformation and
drive the drop towards the equilibrium spherical shape.  Finally
the vorticity parameter $\beta$ characterizes the magnitude of the
rotational component of the external flow relative to the extensional
component.

In this paper we focus on the parameter regime where the drop
deformation may be significant, which requires that
$\Ca=O(1)$. (However, the flow is not strong enough to cause drop
breakup.)  We also assume that the drop-phase fluid is much more
viscous than the continuous-phase fluid,
\begin{equation}
\label{large viscosity ratio}
\lambda\gg1.
\end{equation}
Drop deformation process is hindered at large drop-phase viscosities,
while drop rotation is only weakly affected by the viscous stresses
inside the drops.  Therefore the relative effect of the drop rotation
is amplified in the regime \refeq{large viscosity ratio}, and the
rotational component of the external flow produces nontrivial
qualitative effects.

\begin{figure}
{\includegraphics[width=2.5in]{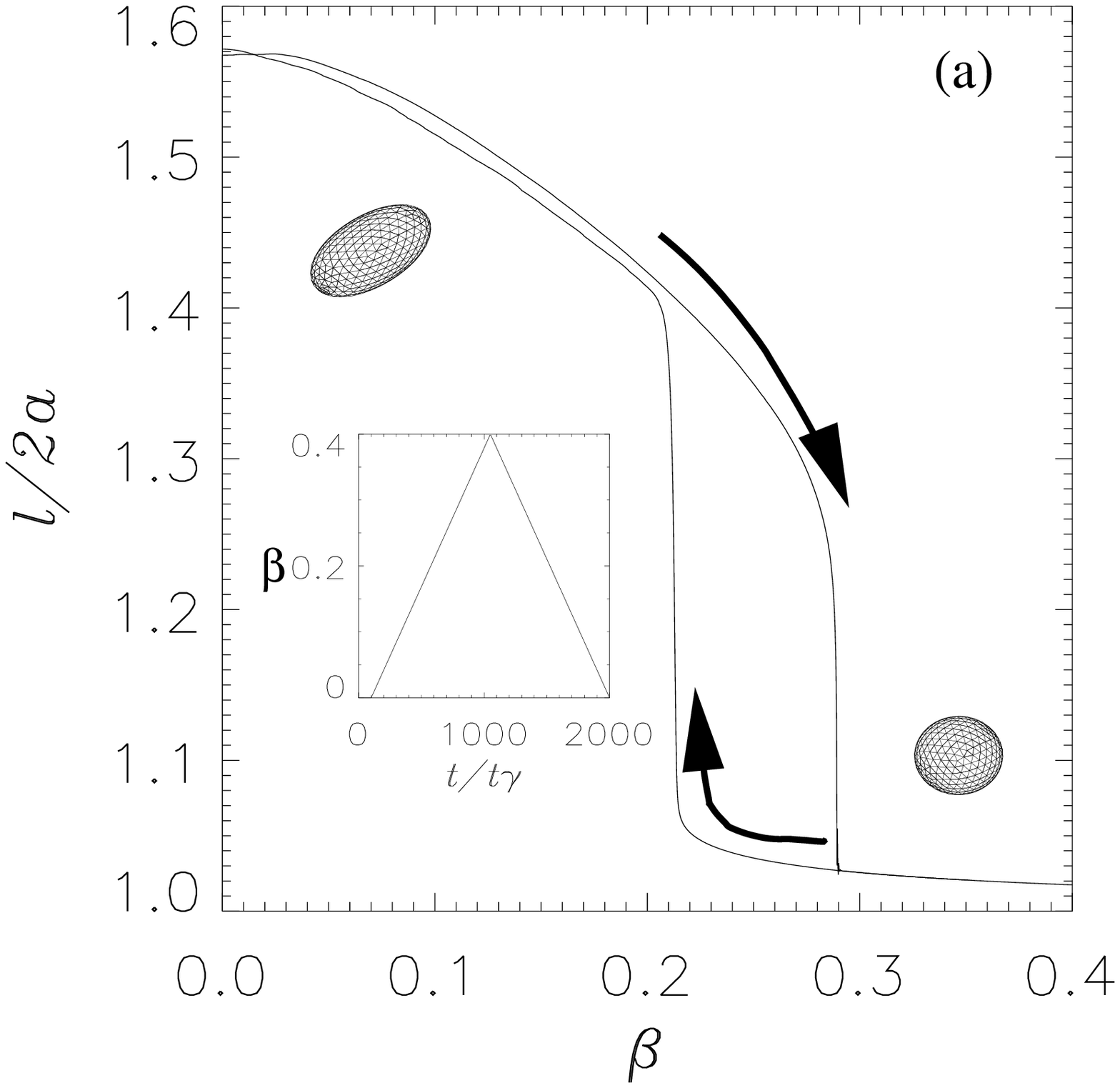}}
{\includegraphics[width=2.5in]{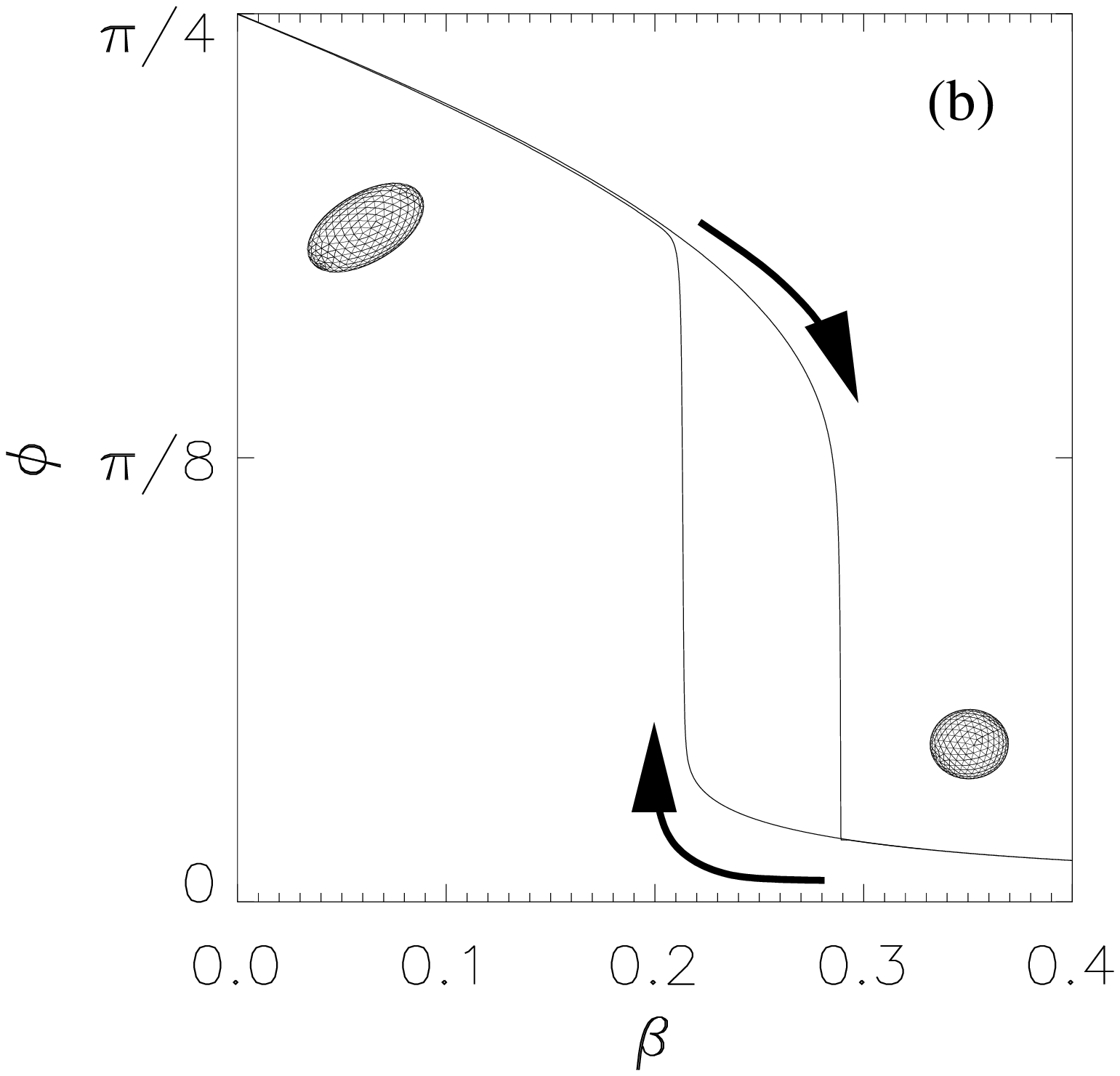}}
\caption{Hysteretic evolution of viscous drop in two-dimensional
straining flow with slowly varying vorticity.  Viscosity ratio
$\lambda=200$ and capillary number $\Ca=0.20$.  \subfig{a} Normalized
drop length $\dropLength$ and \subfig{b} drop angle $\dropAngle$
versus vorticity parameter $\beta$.  Arrows indicate the direction of
increasing time, and inset shows dependence of $\beta$ on time.
(Results from boundary-integral simulations.)
%
%
}
\label{fig:1}
\end{figure}

\section{Hysteretic drop behavior}
\label{Hysteretic behavior}

\subsection{Capillary and rotational stabilizing mechanisms}
\label{Capillary and rotational stabilizing mechanisms}

To illustrate the effect of the rotational component of the flow
\refeq{decomposition of external flow} on the dynamics of a highly
viscous drop we consider a system where the parameter $\beta$ is first
slowly increased and then slowly decreased.  We adopt here a linear
ramping protocol where the vorticity is slowly ramped up from
$\beta=0$ to $\beta=0.4$ and then ramped down back to zero.  Before
the ramping occurs the flow is maintained at $\beta=0$ (for 5\,\% of
the total ramping time) to allow the drop to relax to the stationary
shape in purely straining flow.  This vorticity variation protocol is
represented in the inset of figure \ref{fig:1}\subfig{a}.

The evolution of the drop shape in this time-dependent flow is
depicted in figure \ref{fig:1}.  Figure \ref{fig:1}\subfig{a}
represents the drop length $\dropLength$, and figure
\ref{fig:1}\subfig{b} shows the drop angle $\phi$ (measured
anticlockwise from the axis $x$); both quantities are plotted versus
the vorticity parameter $\beta$.  In our example, the drop viscosity
is $\lambda=200$.  The capillary number $\Ca=0.2$ is below the
critical value for drop breakup ($\Ca=0.22$ in 2d straining flow) but
it is sufficiently large to allow for a significant flow-induced drop
deformation.  The total ramping time is $\Tramp=2000\tauDeformation$,
so that the drop response to the flow
variation is nearly quasistatic.  The calculations were performed
using the adaptive boundary-integral procedure developed by
\cite{Cristini-Blawzdziewicz-Loewenberg:2001,%
Cristini-Blawzdziewicz-Loewenberg:1998a%
}.

The results shown in figure \ref{fig:1} indicate that the drop
response to the vorticity variation is hysteretic.  At $\beta=0$ drop
is elongated, and it is aligned with the extensional axis of the
straining component of the flow ($\phi=\pi/4$).  With increasing
$\beta$, the drop orientation slowly changes towards the symmetry axis
$x$ ($\phi=0$), and the drop length slowly decreases.  At a critical
value of the vorticity parameter, $\betaUp\approx0.29$, a
discontinuous change occurs: the drop length and the angle suddenly
decrease.  Afterwards the drop is almost spherical and nearly aligned
with the axis $x$.  When the direction of the vorticity change is
reversed, the drop initially retraces its trajectory.  However, the
drop does not jump back to the elongated shape until the vorticity
reaches the lower critical value $\betaLow\approx0.22<\betaUp$.

The bistable drop behavior and the associated hysteretic shape
evolution stem from the existence of two mechanisms that can stabilize
a viscous drop in linear flows with rotation \refeq{decomposition of
external flow}.  Namely, the drop can be stabilized by the capillary
stresses (which drive the drop towards the equilibrium spherical
shape) or by the vorticity flow component (which rotates the drop out
of the extensional axis of the straining component of the flow).

In a purely straining flow the drop assumes the
interfacial-tension-stabilized elongated shape \citep{Taylor:1934}.
The form of the drop results from the balance between drop deformation
by the flow and drop relaxation due to the capillary forces.  The
deformation and relaxation occur on the respective time scales
\begin{equation}
\label{deformation time scale}
\tauDeformation=\lambda\dot\gamma^{-1},
\end{equation}
\begin{equation}
\label{relaxation time scale}
\tauRelaxation=\lambda\mu a\sigma^{-1},
\end{equation}
both of which are proportional the viscosity ratio $\lambda\gg1$.  The
drop deformation $D=(\dropLength-2a)/a$ is determined by the time
scale ratio
\begin{equation}
\label{deformation of elongated drop}
D\sim\tauRelaxation/\tauDeformation=\Ca,
\end{equation}
and therefore it is independent of the viscosity ratio in the limit
$\lambda\to\infty$.  In purely staining flows, the drop is oriented
along the extensional axis $x=y$.

\begin{figure}
\vspace{0.2in}
\begin{center}
\includegraphics[width=1.8in]{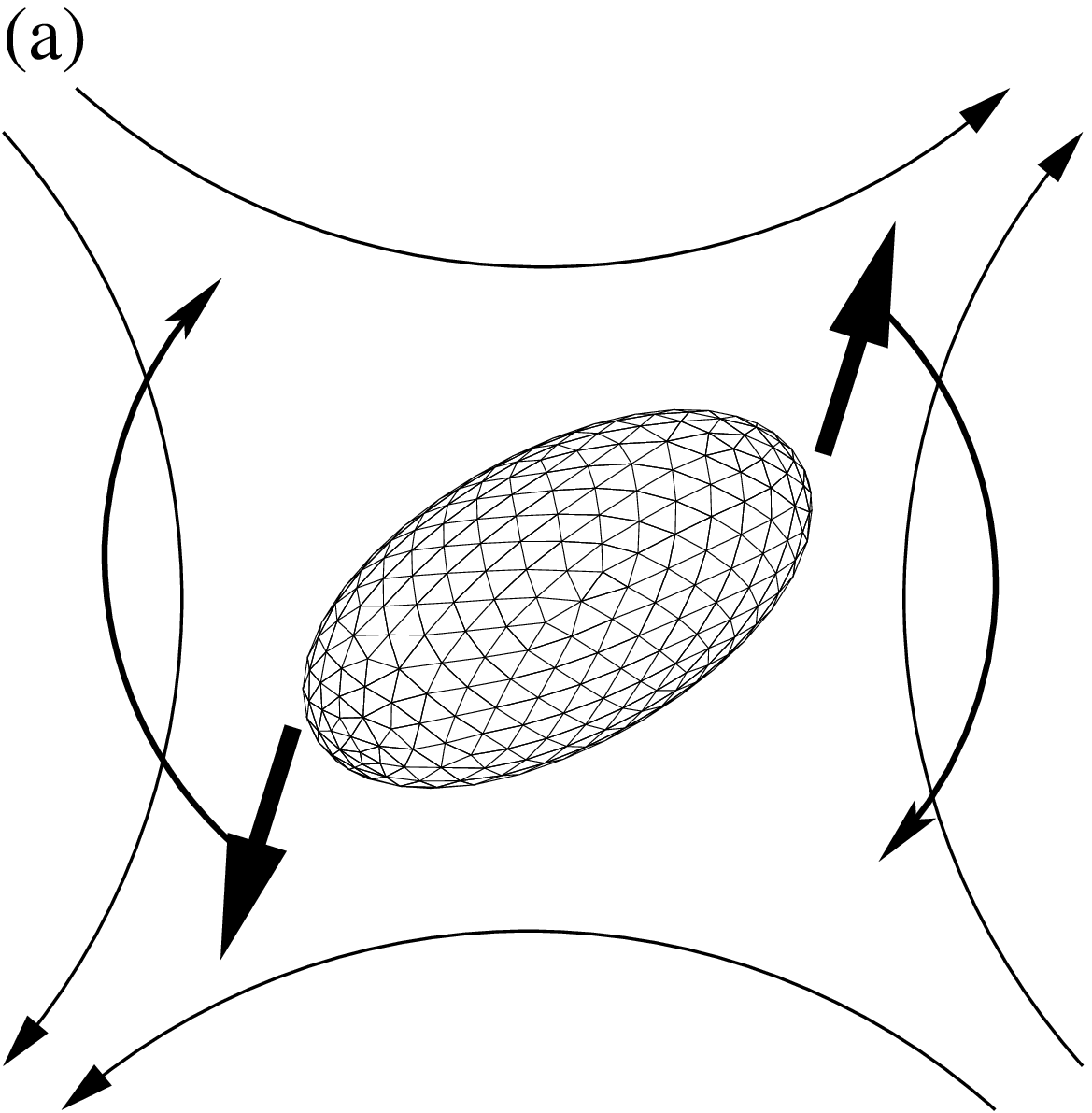}
\hspace{0.8in}
\includegraphics[width=1.8in]{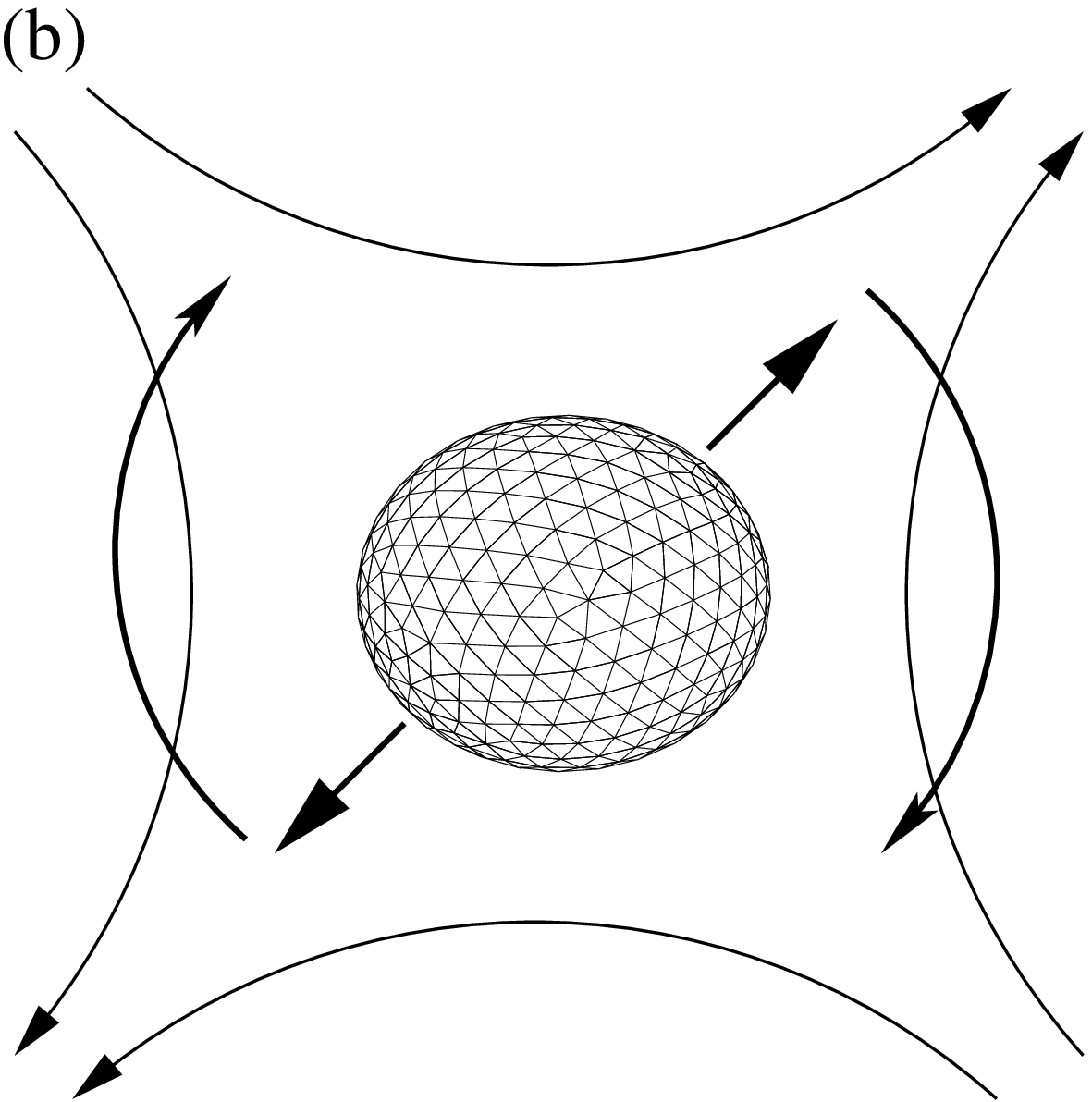}
\end{center}
\caption{Schematic representation of the physical mechanism leading to
bistable drop behavior in two-dimensional linear flows with non-zero
vorticity.  \subfig{a} Elongated drop is stabilized by capillary forces
and destabilized by flow rotation; \subfig{b} compact drop is
stabilized by flow rotation and destabilized by extensional flow
component.}
\label{mechanism}
\end{figure}

For small values of $\beta$, the vorticity flow component produces an
$O(\beta)$ perturbation of the drop orientation.  The corresponding
decrease of the drop length is $O(\beta^2)$.  However, a further drop
rotation is arrested because the straining component of the flow
produces hydrodynamic stresses that pull the elongated drop back
towards the straining axis (as illustrated in  figure
\ref{mechanism}\subfign{a}).

Since the $O(\lambda^{-1})$ internal circulation inside an elongated
high-viscosity drop is weak, the drop in its stationary state behaves
analogously to a rigid body whose equilibrium orientation results from
the balance of the torques produced by the straining and rotational
components of the external flow.  The transition to the compact drop
shape occurs when the vorticity flow component becomes too strong to
be balanced.  Under such conditions, a rigid body would undergo a
transition to a periodic motion with continuous rotation in the
clockwise direction.  Similarly, a drop also starts to continuously
rotate when $\beta$ achieves the upper critical value $\betaUp$.
However, during the rotation the drop length decreases because the
drop becomes misaligned with the extensional axis of the flow.  As a
result, the drop relaxes to a nearly spherical shape.

In this new, compact shape the fluid inside the drop circulates with
the angular velocity $\omegaDrop$ that is nearly equal to the angular
velocity of the external flow \refeq{angular velocity}.  Within each
period of rotation the drop undergoes a small deformation produced by
the straining component of the external flow (as schematically
illustrated in figure \ref{mechanism}\subfign{b}).  However, the
deformation does not accumulate because it is constantly convected
away by the rotational component of the flow.

Since the rotation occurs on the time scale
\begin{equation}
\label{rotation time scale}
\tauRotation=(\beta\dot\gamma)^{-1},
\end{equation}
and the drop deforms on the much longer timescale \refeq{deformation
time scale}, we find that the drop deformation
\begin{equation}
\label{deformation in the compact state}
D\sim\tauRotation/\tauDeformation=(\beta\lambda)^{-1}
\end{equation}
is small for $\lambda\gg1$, consistent with the results shown in
figure \ref{fig:1}\subfig{a}.

Relation \refeq{deformation in the compact state} indicates that the
deformation of the rotationally stabilized drop increases with the
decreasing parameter $\beta$.  When $\beta$ falls below the lower
critical value $\betaLow$, the hydrodynamic torque associated with the
straining component of the flow acting on a slightly elongated drop
becomes strong enough to reorient the drop along the straining axis
and arrest further drop rotation.  Deformation thus starts to
accumulate, the drop is stretched, and a transition to the
interfacial-tension stabilized elongated state takes place.

As shown in figure \ref{fig:1}, a drop in the compact, rotationally
stabilized stationary state is nearly aligned with the symmetry axis
of the applied flow $x$.  This behavior stems from the flow-reflection
symmetry of Stokes equations and the fact that the drop is stabilized
by rotation rather than the capillary forces.  In the absence of the
capillary forces (or in the limit of infinitely strong flow) the
symmetry of Stokes equations implies that the stationary drop shape is
invariant with respect to flow reflection.  Hence, the shape is
also invariant with respect to the corresponding transformation
$(x,y,z)\to(-x,y,z)$ of the spatial coordinates (and this symmetry
corresponds to drop alignment in the $x$ direction).  A perturbation
due to the presence of the capillary stresses produces only a small
asymmetry because the effect of capillary forces is insignificant for
a nearly spherical drop.

\begin{figure}
\begin{center}
{\includegraphics[width=3.2in]{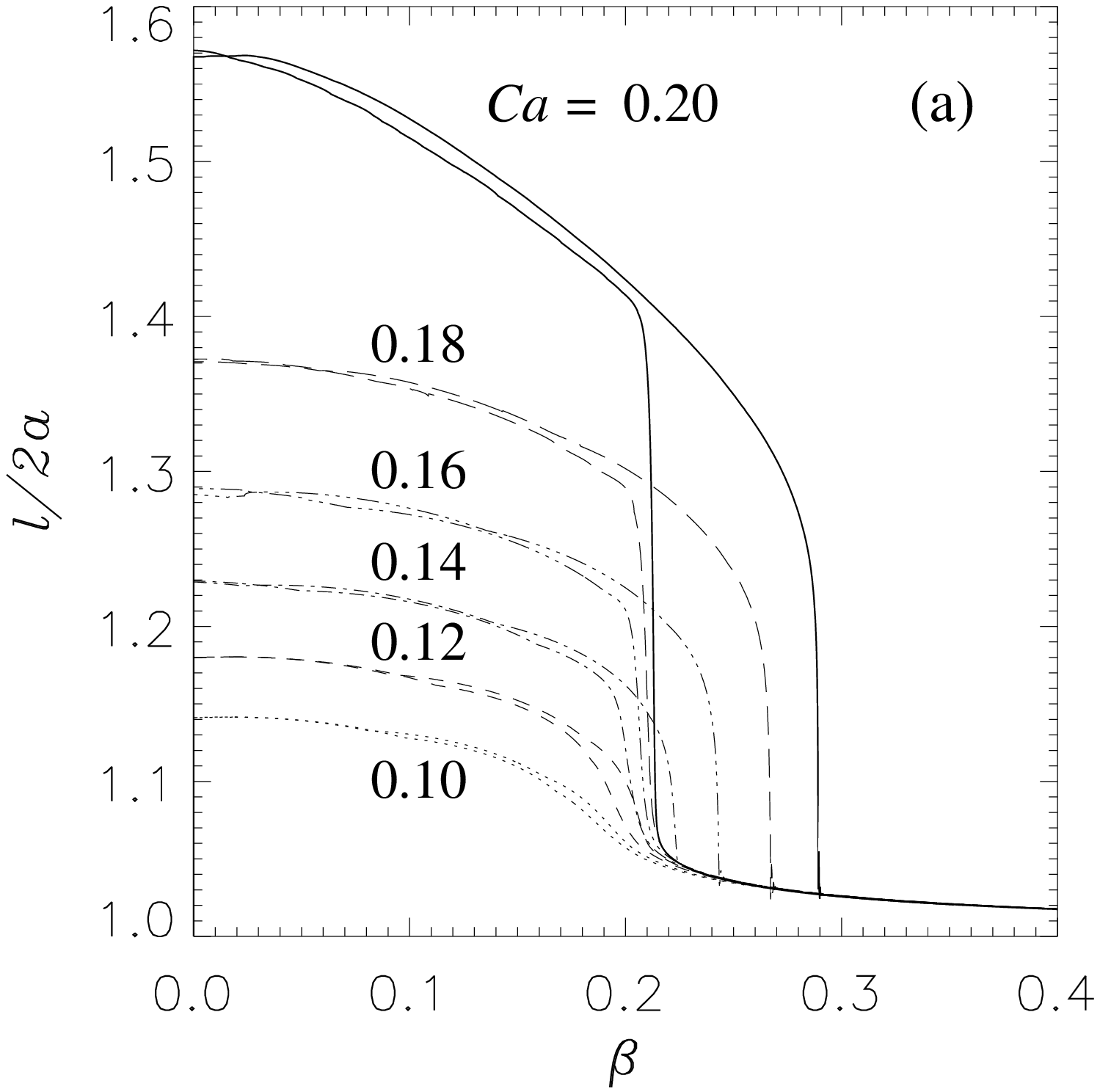}}\\
\end{center}
{\includegraphics[width=2.5in]{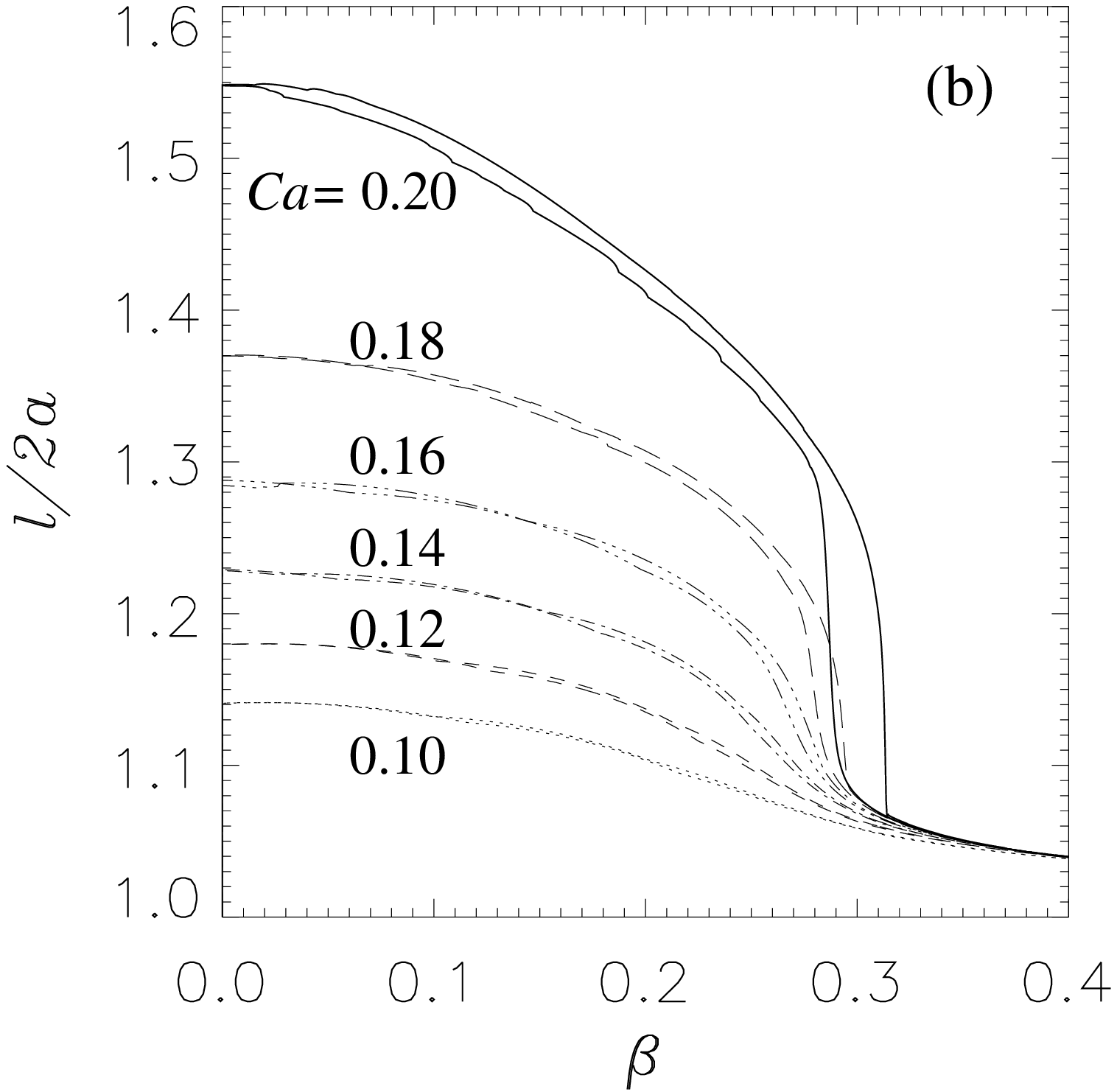}}
{\includegraphics[width=2.5in]{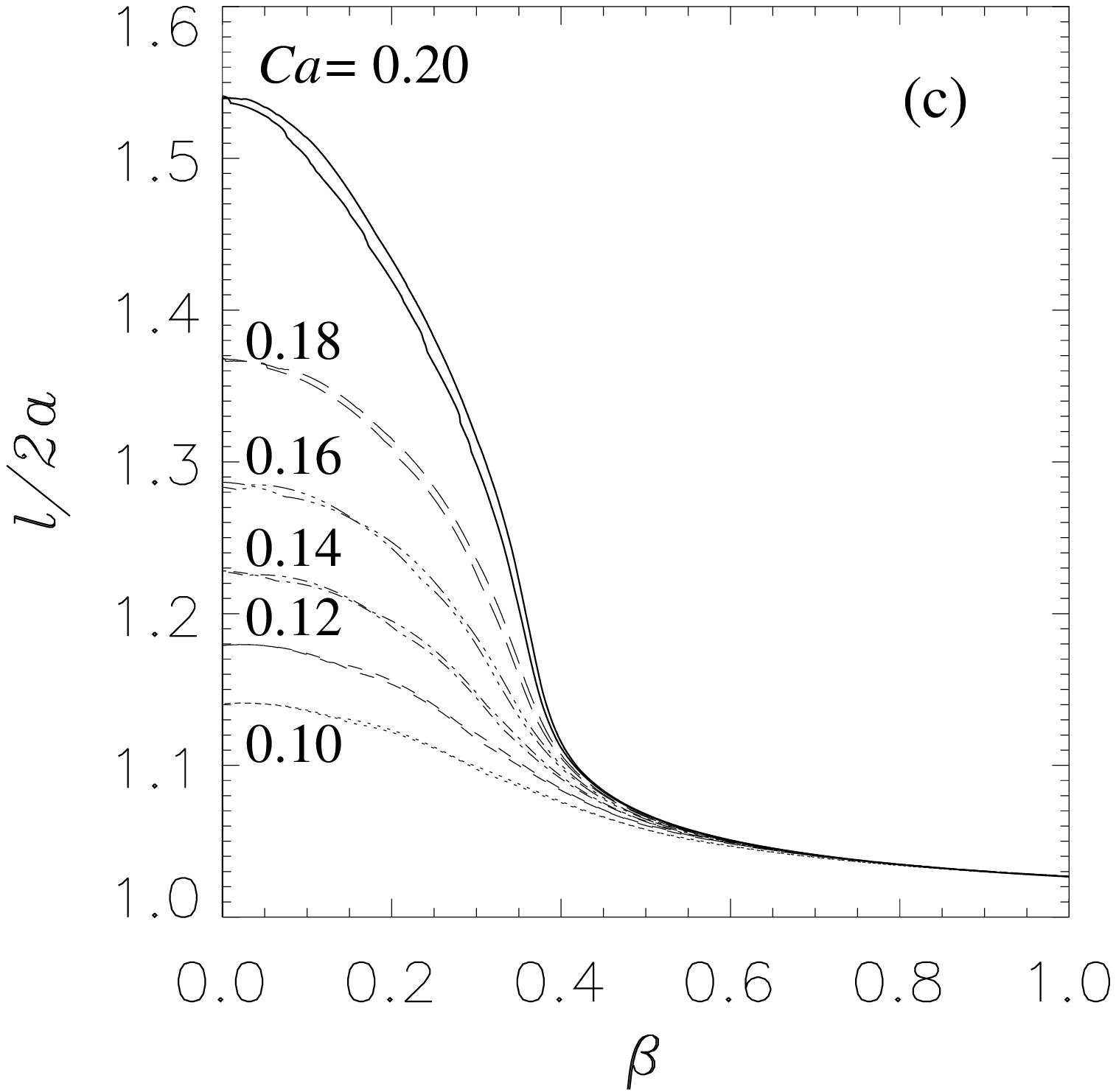}}
\caption{
Quasistatic variation of drop length $\dropLength$ with vorticity
parameter $\beta$ for \subfig{a} $\lambda=200$, \subfig{b} 100,
\subfig{c} 50, and different values of capillary number (as
labeled). (Results from boundary-integral simulations.)  
}
\label{drop response for different Ca and lambda}
\end{figure}

\subsection{Parameter dependence of drop response}
\label{Dependence on system parameters}

The quasistatic response of the drop length to the variation of the
vorticity is illustrated in figure \ref{drop response for different Ca
and lambda} for different values of the capillary number $\Ca$ and
viscosity ratio $\lambda$.  The results indicate that the size of the
hysteresis loop is the largest for large values of $\lambda$ and
$\Ca$.  When the capillary number is decreased, the upper critical
vorticity parameter $\betaUp$ (corresponding to the transition from
the elongated to the compact drop) decreases but the lower critical
parameter $\beta_1$ remains nearly unaffected.  In contrast, the
viscosity ratio $\lambda$ affects primarily the position of the lower
critical parameter $\betaLow$ (corresponding to the transition from
the compact to the elongated drop).

This behavior is consistent with the scaling relations
\refeq{deformation of elongated drop} and \refeq{deformation in the
compact state} and the mechanism of drop bistability explained in
\S\,\ref{Capillary and rotational stabilizing mechanisms}.
According to these relations the deformation of an elongated drop
scales with $\Ca$ and is independent of $\lambda$, while the
deformation of a compact drop is independent of $\Ca$ and scales with
$\lambda^{-1}$.  We recall that the critical states corresponding to
the transitions between the elongated and compact drop shapes
correspond to the points where the maximal torque $\torqueStrain$
exerted by the extensional component of the flow on the drop just
before the transition marginally balances the torque $\torqueRotation$
exerted by the vorticity flow component.  Since $\torqueStrain\sim D$
whereas $\torqueRotation$ is approximately independent of $D$, we
conclude that $\betaLow$ varies with $\lambda^{-1}$ and $\betaUp$ with
the $\Ca$.  This conclusion is consistent with our simulation results.

The plots shown in figure \ref{drop response for different Ca and
lambda} indicate that for a given viscosity ratio, there exists a
critical value of the capillary number $\CaBifurcation$ below which
the drop response to the changes of the flow vorticity does not
exhibit a hysteretic loop.  The bifurcation point occurs at the
critical value of the vorticity parameter $\betaBifurcation$ that
corresponds to the position of the infinitesimal hysteresis loop for
$\Ca$ slightly above the critical value $\CaBifurcation$.  It can be
shown from the results derived by
\cite{Blawzdziewicz-Cristini-Loewenberg:2003} that
\begin{equation}
\label{bifurcation parameters}
\CaBifurcation=\frac{15}{19}\left(\frac{5}{\lambda}\right)^{1/2},
\qquad
\betaBifurcation=\frac{3}{4}\left(\frac{15}{\lambda}\right)^{1/2}
\end{equation}
for $\lambda\gg1$.  The plots shown in figure \ref{drop response for
different Ca and lambda} are consistent with the above expressions.
We note that for $\lambda=50$ (see figure \ref{drop response for
different Ca and lambda}\subfign{c}) the hysteretic drop behavior is
not observed.  The reason is that in this viscosity range
$\CaBifurcation$ exceeds the critical value of the capillary parameter
for drop breakup.  Hence, the elongated stationary drop shape does not
exist for $\Ca>\CaBifurcation$.

\subsection{Small-deformation analysis}
\label{Small-deformation analysis}

\subsubsection{Evolution equations}
\label{Small-deformation expansion}

Crucial features of the evolution of a highly viscous drop in
two-dimensional flows with nonzero vorticity are captured by
small-deformation equations.  In our approach
\citep{Blawzdziewicz-Cristini-Loewenberg:2003}, the
position $\rSurface$ of the drop interface is expanded in spherical
harmonics

\begin{equation}
\label{drop interface}
\rSurface/a=\alpha'+\sqrt{2}\sum_{l,m}[f'_{lm} \Real(Y_{lm})
                                     +f''_{lm}\Imaginary(Y_{lm})],
\end{equation}
where $l>0$ and $l\ge m\ge 0$ denote the spherical-harmonic order,
$f'_{lm}$ and $f''_{lm}$ are the expansion coefficients, and the
parameter $\alpha'$ is given by the drop-volume constraint.  Since for
$m=0$ all spherical harmonics are real, we set $f''_{l0}=0$ in
expansion \refeq{drop interface}.  Moreover, flow-induced drop
deformation preserves the symmetry of the incident flow
\refeq{decomposition of external flow}.  Therefore only even
values of $l$ and $m$ need to be included in the analysis.

Evolution equation for the expansion coefficients $f'_{lm}$ and
$f''_{lm}$ are obtained by inserting the series \refeq{drop interface}
into the boundary-value problem \refeq{eq:stokes01}--\refeq{strain and
vorticity tensors}, performing a boundary-perturbation analysis, and
reexpanding resulting products of spherical harmonics using
appropriate Clebsch--Gordan coupling coefficients.  The detailed
analysis and explicit expressions for the evolution equations at
different truncation levels are presented elsewhere
\citep{Vlahovska:2003,Vlahovska-Loewenberg-Blawzdziewicz:2005}.

For simplicity, our small-deformation calculations are performed with
the expansion \refeq{drop interface} truncated at the lowest
spherical-harmonic order $l=2$, which leaves us with three independent
drop-shape components: $f'_{22}, f''_{22}$, and $f'_{20}$.  Noting
that
\begin{equation}
\label{form of spherical harmonics}
\Real(Y_{22})\sim\cos2\phi,\qquad \Imaginary(Y_{22})\sim\sin2\phi,
\end{equation}
we find that the shape parameters $f'_{22}$ and $f''_{22}$ correspond
to the drop deformation along the symmetry axis $x$ and the straining
axis $x=y$, respectively.  The parameter $f'_{20}$ described an
axisymmetric deformation along the axis $z$.

The evolution equations for the shape parameters $f'_{22}, f''_{22}$,
and $f'_{20}$, truncated at the second-order in the drop deformation,
can be represented in the following form
\begin{subequations}
\label{small-deformation equations}
\begin{eqnarray}
\label{equation f20}
\dot{f'_{20}} &=& \lambda^{-1}(d_{11}
   +d_{12}f'_{20})f''_{22}-\lambda^{-1}\Ca^{-1}
\left[D_1f'_{20}-D_2(f^{\prime\hspace{0.5pt}2}_{20}
   -f^{\prime\hspace{0.5pt}2}_{22}-f^{\prime\prime\hspace{0.5pt}2}_{22}
                                                            )\right],\\
\label{equation f'22}
\dot{f'_{22}} &=& -2\omega f''_{22}+\lambda^{-1}\left[d_{21}f'_{22}f''_{22}
    -\Ca^{-1}(D_1+2D_2f'_{20})f'_{22}\right],\\
\label{equation f''22}
\dot{f''_{22}} &=&  2\omega f'_{22}+\lambda^{-1}
    \left[(d_{31}+d_{32}f'_{20}
      +d_{33}f^{\prime\hspace{0.5pt}2}_{20}
      +d_{34}f^{\prime\hspace{0.5pt}2}_{22}
      +d_{35}f^{\prime\hspace{0.5pt}2}_{22})-
                  \Ca^{-1}(D_1+2D_2f'_{20})f''_{22}\right],\nonumber\\
\end{eqnarray}
\end{subequations}
where the dot denotes the time derivative (normalized by
$\dot\gamma^{-1}$).  The terms involving the coefficients $d_{ij}$
correspond to drop deformation by the external flow, and the terms
involving $D_k$ describe the capillary relaxation.  All these terms
are $O(\lambda^{-1})$ in the large-viscosity-ratio regime.  Explicit
expressions for the coefficients $d_{ij}$ and $D_k$ are given in
\citep{Vlahovska:2003}; here we only note that these coefficients
are functions of the viscosity ratio $\lambda$ and have finite limits
for $\lambda\to\infty$.

The two remaining terms on the right-hand side of equations
\refeq{equation f'22} and \refeq{equation f''22} (the terms
proportional to $\omega$) are viscosity independent.  These terms
represent the rigid-body rotation of the drop, with the angular
velocity
\begin{equation}
\label{angular velocity in small deformation equations}
\omega=-\halff\beta+\halff c_1f'_{22},
\end{equation}
where $c_1=(15/2\pi)^{1/2}$.  Consistent with our
qualitative physical picture described in \S\,\ref{Capillary and
rotational stabilizing mechanisms} (and illustrated in figure
\ref{mechanism}), the rotational velocity \refeq{angular velocity in
small deformation equations} involves two terms.  The first term
corresponds to the rotation of the drop by the vorticity component of
the flow \refeq{decomposition of external flow}.  The second term,
which is proportional to the shape parameter $f'_{22}$ that described
deformation in the $x$ direction, corresponds to the rotation of a
deformed drop by the straining component of the external flow towards
the straining axis $x=y$.

\subsubsection{Reduced description}
\label{Reduced description}

It has been shown by \cite{Blawzdziewicz-Cristini-Loewenberg:2003}
that for $\lambda\gg1$ the drop behavior near the bifurcation point
\refeq{bifurcation parameters} can be described by simplified
asymptotic equations
\begin{subequations}
\label{simplified evolution equation}
\begin{eqnarray}
\label{simplified evolution equation f'22}
\dot f'_{22} &=&  -2\omega f''_{22}
                   -\lambda^{-1}\Ca^{-1}\Dcoefficient f'_{22},\\
\label{simplified evolution equation f''22}
\dot f''_{22} &=& \hphantom{-}2\omega f'_{22}
                  -\lambda^{-1}\Ca^{-1}\Dcoefficient f''_{22} 
                  +\lambda^{-1}\dCoefficient.
\end{eqnarray}
\end{subequations}
where $\Dcoefficient=20/19$ and $\dCoefficient=(5\pi/6)^{1/2}$
are the high-viscosity
limits of $D_1$ and $d_{31}$.  The asymptotic result \refeq{simplified
evolution equation} is obtained form \refeq{small-deformation
equations} on assumption that near the bifurcation point there is a
balance between drop deformation and rotation (which corresponds to
$\lambda^{-1}\sim\omega f'_{22}$) and the balance between capillary
relaxation and rotation (which yields $(\lambda\Ca)^{-1}f'_{22}\sim
f''_{22}$ and $(\lambda\Ca)^{-1}f''_{22}\sim f'_{22}$).  Moreover, the
two contributions to the angular velocity \refeq{angular velocity in
small deformation equations} are of the same order but do not cancel
(i.e., $\omega\sim\beta\sim f'_{22}$).  Equations
\refeq{small-deformation equations} are rescaled accordingly, and only
the leading order terms are retained.

Equations \refeq{simplified evolution equation} have all necessary
ingredients that are needed to describe the hysteretic drop behavior.
We have terms representing drop rotation by the straining and
vorticity components of the external flow (i.e., the terms
proportional to $\omega$), drop relaxation due to the presence of the
capillary forces (the terms proportional to $\Ca^{-1}$), and
stretching of the drop along the straining axis of the straining
component of the external flow (the last term in equation
\refeq{simplified evolution equation f''22}).  Neglecting any of these
terms would qualitatively alter the solution structure, which no longer
would manifest the key features of the drop evolution in the parameter
range considered herein.

\subsubsection{Asymptotic solution}
\label{Asymptotic solution}

In the regime $\lambda^{-1}\ll1$ and $\beta\ll1$ the stationary
solutions of equations \refeq{simplified evolution equation} can be
obtained by a singular-perturbation analysis.  To the leading order in
the small parameters, we find that the elongated drop is described by
the relations
\begin{subequations}
\label{equations for elongated drop}
\begin{equation}
\label{equation 1 for elongated drop}
\omega\simeq0,
\end{equation}
\begin{equation}
\label{equation 2 for elongated drop}
\lambda^{-1}\Ca^{-1}\Dcoefficient f''_{22}\simeq\lambda^{-1}\dCoefficient.
\end{equation}
\end{subequations}
The first of the above relations correspond to the fact that an
elongated drop does not rotate (there is only a weak fluid circulation
inside it, as predicted by our qualitative analysis).  The second
relation describes the balance between drop deformation by the
external flow and relaxation due to the capillary forces.  Recalling
the definition \refeq{angular velocity in small deformation equations}
of the angular velocity $\omega$ we find that relations
\refeq{equations for elongated drop} yield
\begin{subequations}
\label{solution for elongated drop}
\begin{equation}
\label{f' for elongated drop}
f'_{22}\simeq c_1\beta,
\end{equation}
\begin{equation}
\label{f'' for elongated drop}
\quad f''_{22}\simeq\Dcoefficient^{-1}\dCoefficient\,\Ca.
\end{equation}
\end{subequations}
By inserting the above relations back into \refeq{simplified evolution
equation} one can verify that they constitute a consistent
leading-order asymptotic stationary solution.

According to equation \refeq{f'' for elongated drop} drop elongation
along the straining axis $x=y$ (i.e., $\phi=\pi/4$) scales with the
capillary number, which is consistent with the scaling result
\refeq{deformation of elongated drop}.  The drop angle 
\begin{equation}
\label{drop angle}
\phi=\halff\arctan(f''_{22}/f'_{22})
\end{equation}
only slightly deviates from $\phi=\pi/4$ because $f'_{22}\ll f''_{22}$
(assuming that $\Ca\gg\beta$ and $\beta\ll1$).

The leading-order stationary solution corresponding to the compact
drop is obtained from the following relations
\begin{subequations}
\label{equations for compact drop}
\begin{equation}
\label{equation 1 for compact drop}
-2\omega f''_{22}\simeq0,
\end{equation}
\begin{equation}
\label{equation 2 for compact drop}
2\omega f'_{22}\simeq-\lambda^{-1}\dCoefficient, 
\end{equation}
\end{subequations}
which are obtained by dropping from the evolution equations
\refeq{simplified evolution equation} the $O(\lambda^{-1})$
capillary-relaxation terms.  Taking into account the definition
\refeq{angular velocity in small deformation equations} of $\omega$ we
thus obtain 
\begin{subequations}
\label{solution for compact drop}
\begin{equation}
\label{f'' for compact drop}
\quad f''_{22}\simeq0
\end{equation}
and
\begin{equation}
\label{f' for compact drop}
f'_{22}\simeq\frac{\beta-\sqrt{\beta^2-4c_1\dCoefficient\lambda^{-1}}}{2c_1}
\end{equation}
\end{subequations}
(the solution with the plus sign in front of the square root is
unstable).  

Since the shape parameter $f''_{22}$ vanishes according to equation
\refeq{f'' for compact drop}, the drop is oriented in the $x$
direction.  For $\lambda^{-1}\ll\beta$ we find
\begin{equation}
\label{expansion of short drop}
f'_{22}=\dCoefficient\beta\lambda^{-1},
\end{equation}
which is consistent with our scaling estimate \refeq{deformation in
the compact state}.  For $\beta<\beta_1$, where
\begin{equation}
\label{expression for beta low}
\beta_1=2(c_1\dCoefficient)^{1/2}\lambda^{-1/2},
\end{equation}
the solution \refeq{f' for compact drop} does not exist; the drop thus
undergoes a transition to the long solution.  The critical vorticity
parameter decreases with increasing $\lambda$, in agreement with our
numerical results presented in figure \ref{drop response for different
Ca and lambda}.

The solution \refeq{equations for elongated drop}--\refeq{expression
for beta low} of the simplified small-deformation equations
\refeq{simplified evolution equation} is perturbative.  We note,
however, the exact stationary solution can also be found
\citep{Blawzdziewicz-Cristini-Loewenberg:2003}.  Our analytical
solution quantitatively agrees with the results of numerical
simulations, provided that the drop deformation is not too large.

\begin{figure}
\begin{center}
{\includegraphics[width=2.5in,height=2.7in]{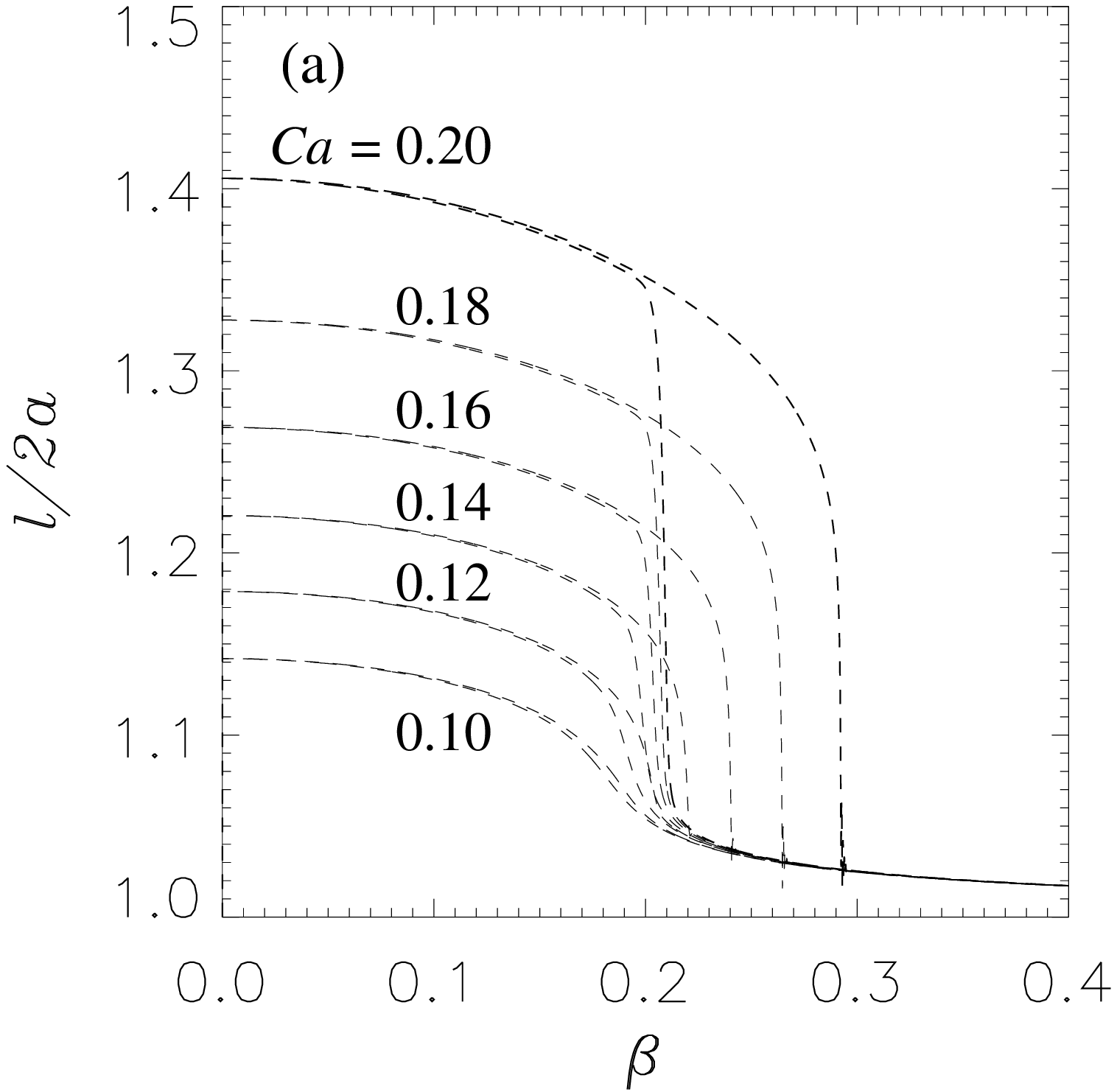}}
{\includegraphics[width=2.5in,height=2.7in]{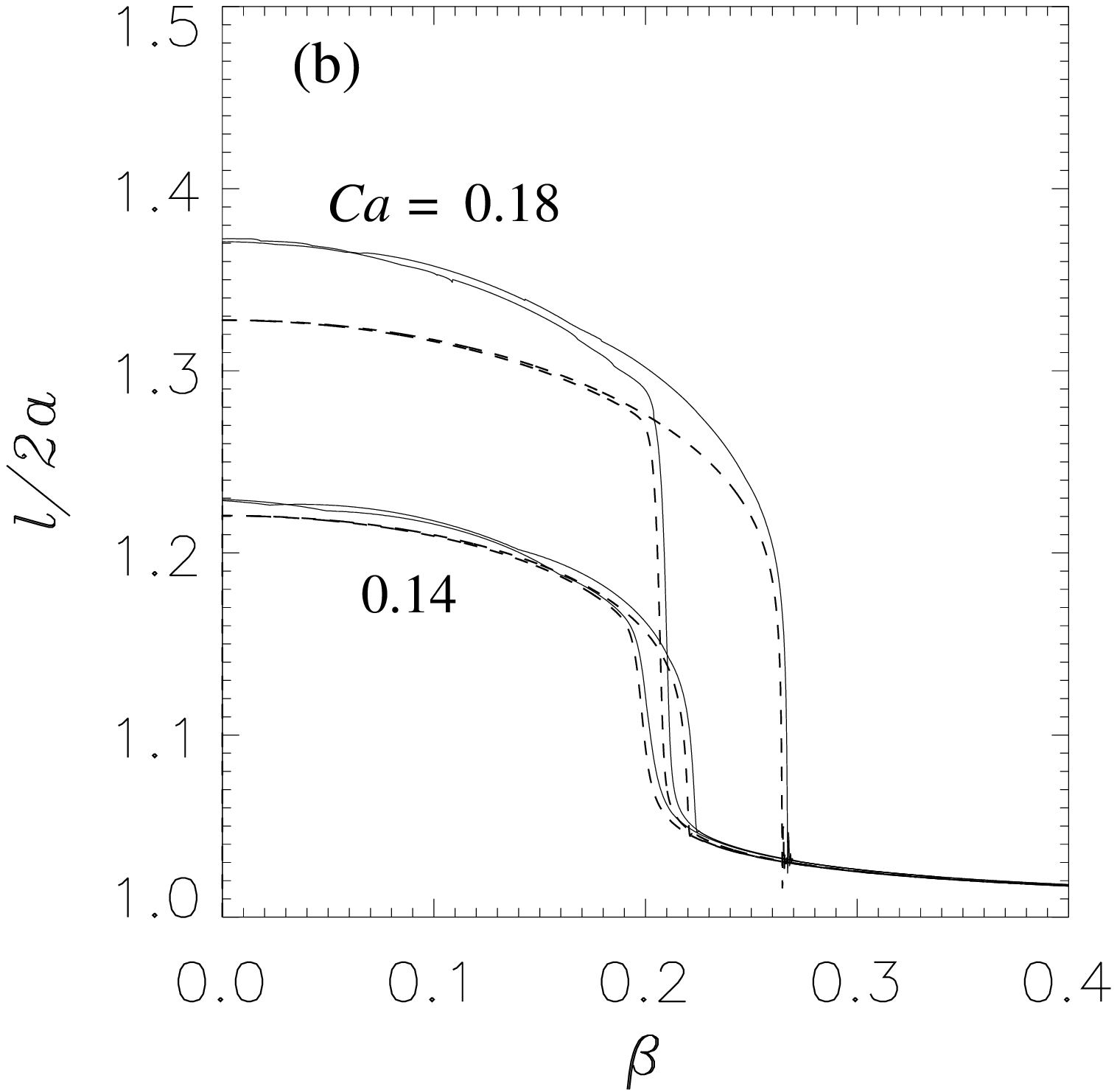}}
\end{center}
\caption{ 
Quasistatic variation of drop length $\dropLength$ with vorticity
parameter $\beta$ for $\lambda=200$ and different values of capillary
number (as labeled).  \subfig{a} Solution of small-deformation
equations (\protect\ref{small-deformation equations}); \subfig{b}
comparison of small-deformation results (dashed lines) with boundary
integral simulations (solid lines). 
}
\label{small deformation results for lambda=200}
\end{figure}

\subsubsection{Numerical results}
\label{Comparison with numerical results}

Predictions of the small-deformation equations
\refeq{small-deformation equations} for drop behavior in a
two-dimensional linear flow with slowly varying vorticity are depicted
in figure \ref{small deformation results for lambda=200} for a system
with the viscosity ratio $\lambda=200$.  Figure \ref{small deformation
results for lambda=200}\subfig{a} shows the dependence of the drop
length
\begin{equation}
\label{drop lenght from small deformation}
\dropLength/a=1+\sqrt{\frac{15}{8\pi}}
  (f^{\prime\hspace{0.5pt}2}_{22}+f^{\prime\prime\hspace{0.5pt}2}_{22})^{1/2}
\end{equation}
on the vorticity parameter $\beta$ for the same set of capillary
numbers as those represented in figure \ref{drop response for
different Ca and lambda}.  Figure \ref{small deformation results for
lambda=200}\subfig{b} compares the small deformation results directly
with the results of the boundary-integral simulations.  The
small-deformation calculations were performed using the second-order
equations \refeq{small-deformation equations} because for an elongated
drop they are more accurate than the simplified equations
\refeq{equations for elongated drop}.  

The results shown in figures \ref{drop response for different Ca and
lambda} and \ref{small deformation results for lambda=200} indicate
that for small and moderate capillary numbers the small-deformation
theory yields accurate quantitative predictions.  At high values of $\Ca$
drop behavior is also captured quantitatively, except for the upper
portion of the hysteresis loop (i.e.\ when the drop is in the
elongated state).  For all values of the capillary number the lower
and upper critical vorticity parameters $\beta_1$ and $\beta_2$ are
obtained within the numerical error of the boundary-integral
simulations.  Our additional calculations (not shown) indicate that a
similar accuracy is obtained for $\lambda=50$ and $\lambda=100$.

\begin{figure}
$\begin{array}{lc}
{\includegraphics[width=2.5in,height=4.8in]
{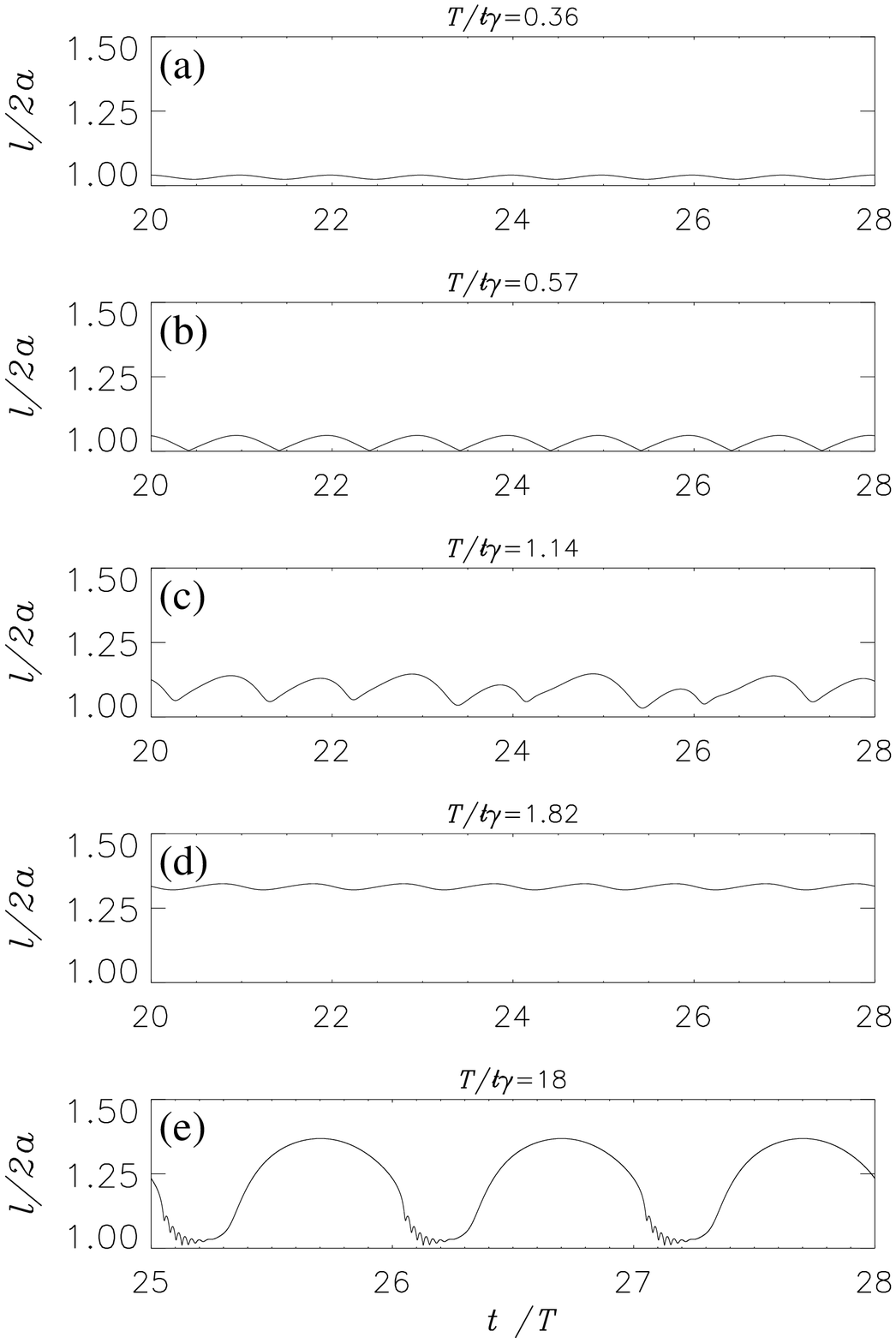} } &
{\includegraphics[width=2.5in,height=4.8in,trim=20 0 0 0]
{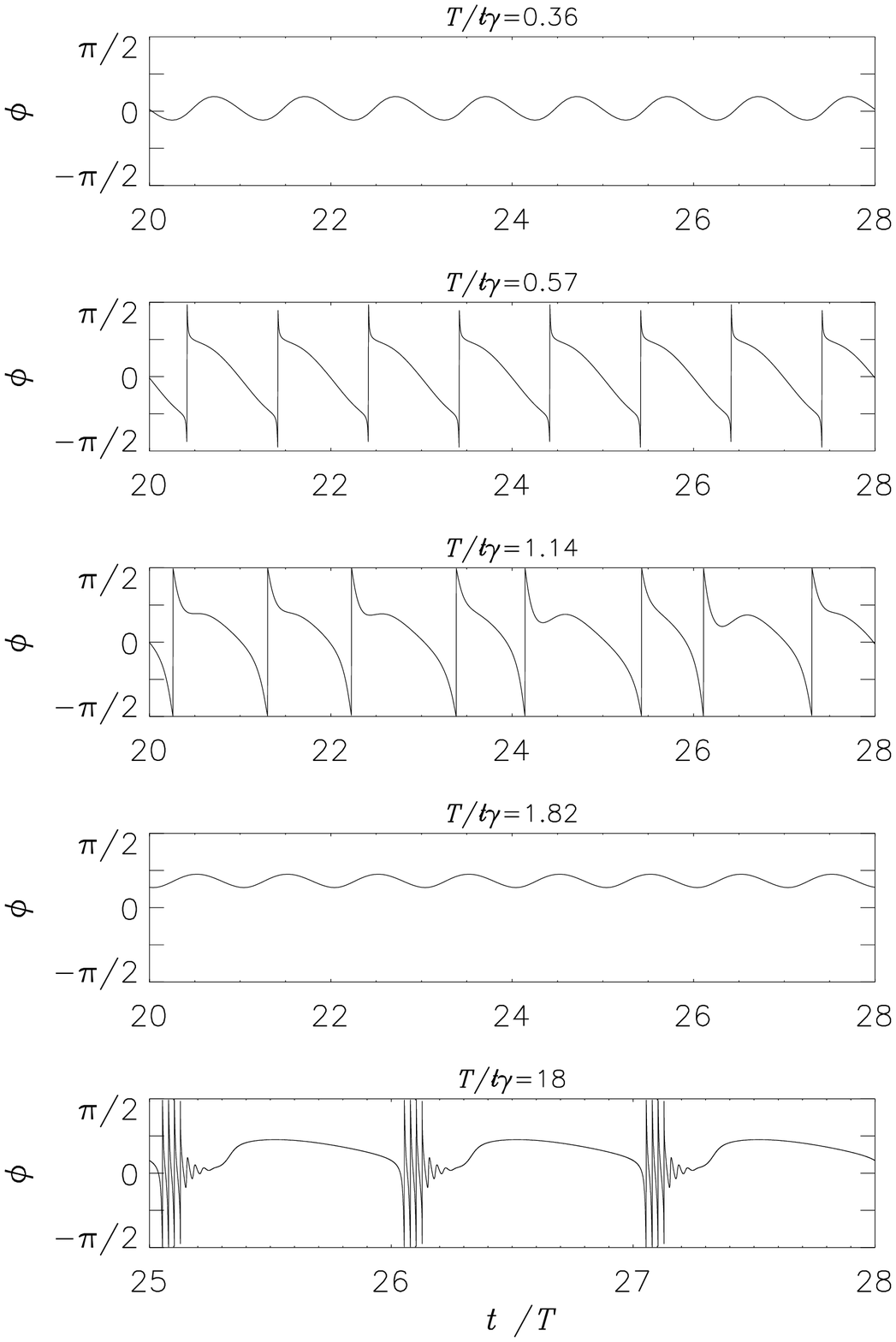} }
\end{array}
$
\caption{
Evolution of drop length (left) and angle (right) in two dimensional
linear flow with harmonic variation of vorticity
(\protect\ref{sinusoidal vorticity variation}), for different values
of period $T$ normalized by drop-deformation time (as labeled).  Mean
vorticity $\betaMean=0.25$, vorticity amplitude $\betaVariance=0.13$,
viscosity ratio $\lambda=275$ and capillary number $\Ca=0.2$.  Panels
\subfig{c} depict chaotic dynamics. (Results from mall-deformation
theory.)
}
\label{fig:chaos01}
\end{figure}
\begin{figure}
\begin{center}
\includegraphics[width=4.5in]
{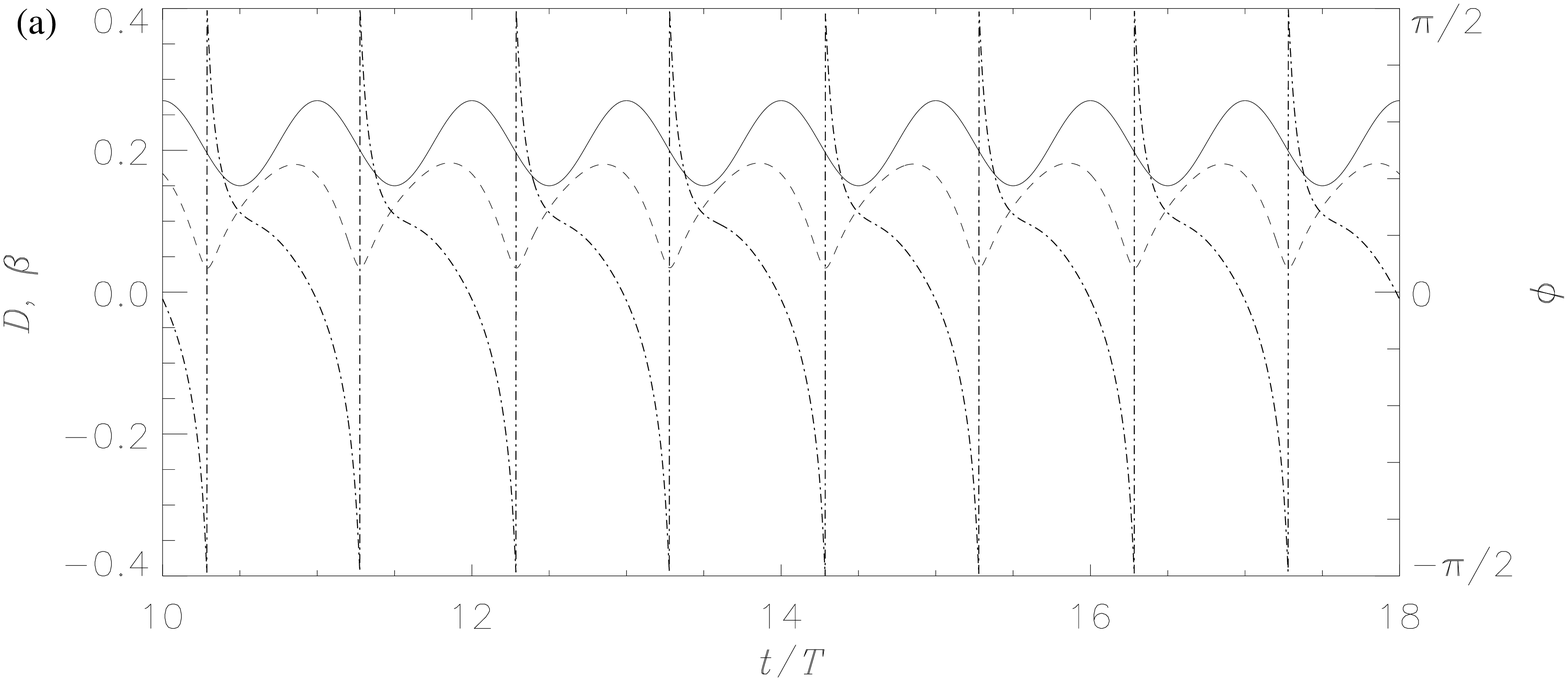} 
\includegraphics[width=4.5in]
{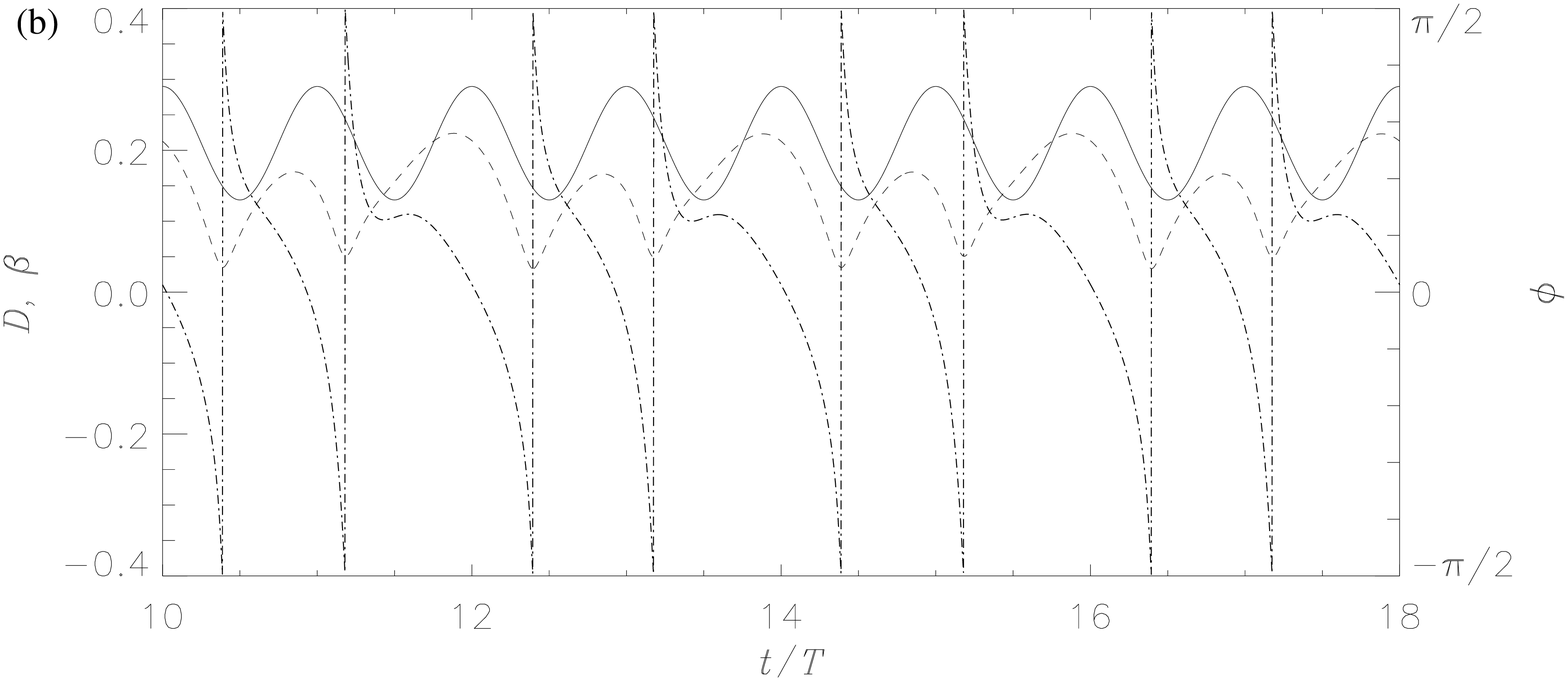} 
\end{center}
\caption{Period doubling in the dynamics of viscous drop in
two-dimensional linear flow with harmonic variation of vorticity.
Vorticity parameter $\beta$ (solid lines), drop deformation $D$
(dashed), and drop angle $\dropAngle$ (dash-dotted) are shown versus
time $t$ normalized by the oscillation period $T$.  Viscosity ratio
$\lambda=275$, capillary number $\Ca=0.2$, period $T/\tauDeformation=1.14$,
mean vorticity $\betaMean=0.21$, and vorticity oscillation
amplitude \subfig{a} $\deltaBeta=0.6$ and \subfig{b} 0.8.  (Results
from small-deformation theory.)
}
\label{example of period doubling}
\end{figure}

\section{Chaotic drop dynamics in a sinusoidal straining flow}
\label{sec:sinusoidal_chaos}

Dynamical systems with multiple equilibrium states often exhibit novel
dynamics when driven by simple forcing \citep{Guckenheimer}.  Thus,
despite the laminar nature of the Stokes flow, we expect to find
interesting nonlinear dynamics of a viscous drop in a time-varying
linear flow with rotation.  To explore this dynamics we will now
investigate the drop response to harmonic variation of the vorticity
\begin{equation}
\label{sinusoidal vorticity variation}
\beta(t)=\betaMean+\deltaBeta\cos(2\pi t/T),
\end{equation}
where $\betaMean$ is the average vorticity value, $\deltaBeta$ is the
oscillation amplitude, and $T$ is the oscillation period.

We have performed a series of small deformation calculations
(\S\,\ref{subsec:chaos_sdt}) and boundary-integral simulations (\S\,\ref{subsec:chaos_bis}) for different values of the flow parameters
$\betaMean$, $\deltaBeta$, and $T$.  If the oscillation period $T$ is
much shorter than the drop deformation and oscillation times
\refeq{deformation time scale} and \refeq{relaxation time scale}, we
find that the drop undergoes small oscillations about a stationary
shape corresponding to the mean value of $\beta$ (which is an expected
behavior).  In the opposite regime
$T\gg\tauDeformation,\tauRelaxation$, the quasistatic drop behavior
described in \S\,\ref{Hysteretic behavior} is recovered.  In what
follows we focus on the most interesting parameter domain
$T\sim\tauDeformation,\tauRelaxation$ and
$\deltaBeta\sim\beta_1-\beta_2$, in which an interaction of different
timescales as well as an interplay between the short and elongated
drop shapes is anticipated.

\subsection{Small-deformation results}
\label{subsec:chaos_sdt}

\begin{figure}
\begin{center}
  {\includegraphics[width=3.5in]{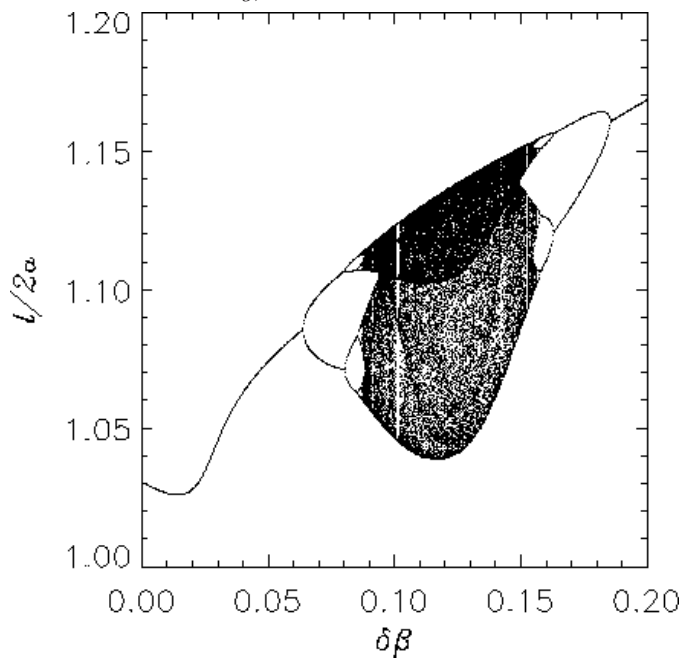} }
\end{center}
\caption{Bifurcation diagram showing period doubling cascades and
transition to chaos for viscous drop in linear flow with harmonic
vorticity variation with mean $\betaMean=0.21$ and period
$T/\tauDeformation=1.14$. Viscosity ratio and capillary number are the
same as in figures \protect\ref{fig:chaos01} and \protect\ref{example
of period doubling}.  (Results from small-deformation theory.)  }
\label{fig:period_doubling}
\end{figure}

Figure \ref{fig:chaos01} illustrates the dependence of the drop
evolution in linear flow with the oscillatory vorticity
\refeq{sinusoidal vorticity variation} on the oscillation period $T$.
The viscosity ratio in this example is $\lambda=275$, and the
capillary number is $\Ca=0.2$.  The mean value of the vorticity
$\betaMean=0.21$ is close to the lower critical value
$\beta_1=0.18$, and the oscillation amplitude is
$\deltaBeta=0.13$.  

Figure  \ref{fig:chaos01}\subfig{a} represent our results for
the shortest oscillation period of the flow vorticity
$T/\tauDeformation=0.36$.  The drop oscillates about the compact
stationary shape in this case.  Both the drop lengths and the drop
angle vary periodically, with the period $T_\drop$ equal to the period
$T$ of the external forcing.  The drop length decreases when the drop
is in the compressional quadrant $-\pi/2<\dropAngle<0$ and increases
for $0<\dropAngle<\pi/2$.  Upon an increase of the period of the
external forcing the amplitude of the angular drop oscillations
increases.  When the oscillation amplitude reaches $\pi/2$ the drop
starts to tumble, as illustrated in figure
\ref{fig:chaos01}\subfig{b}.

\begin{figure}
$\begin{array}{lc}
{\includegraphics[width=2.5in,height=4.8in]
{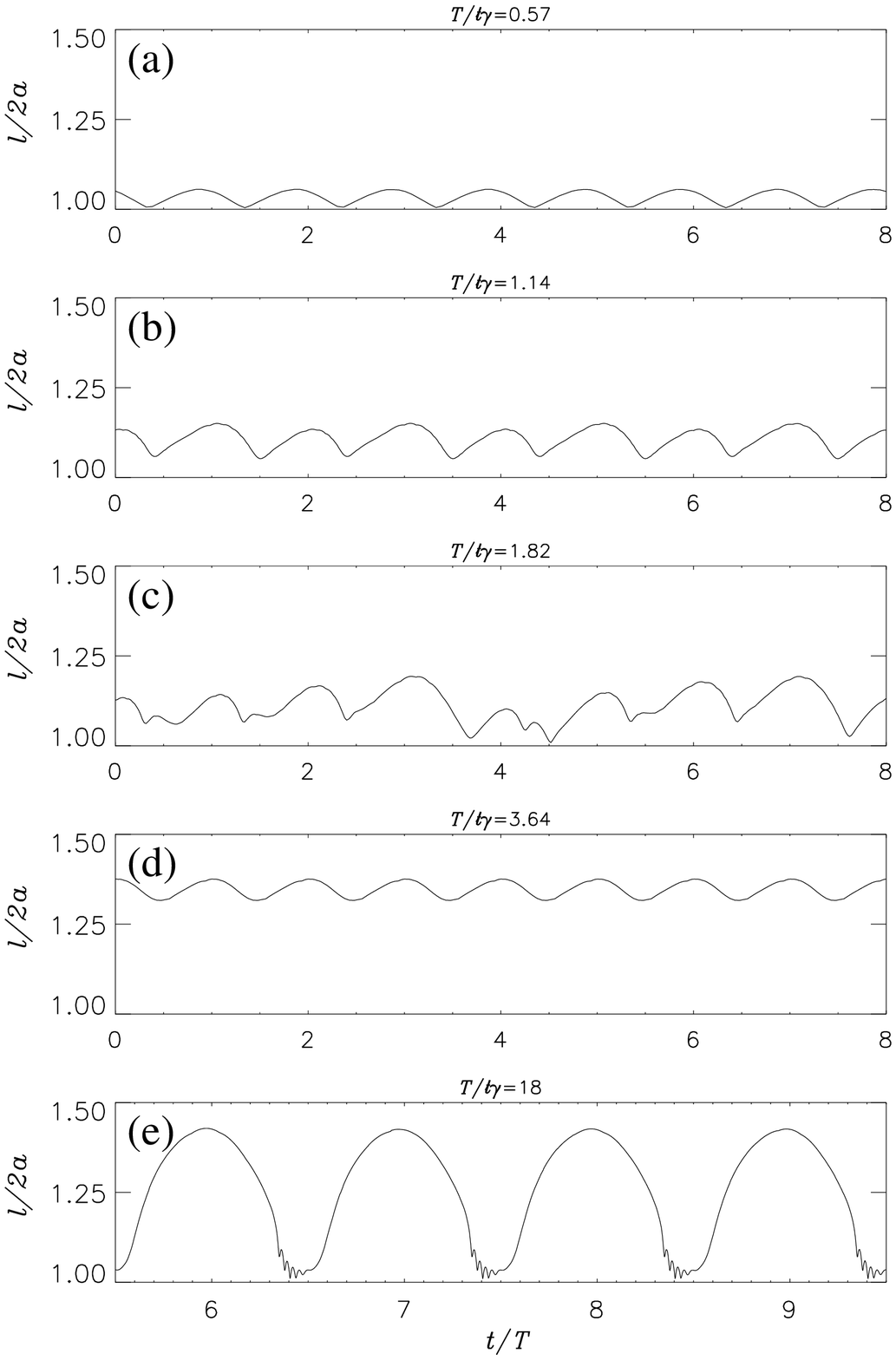} } &
{\includegraphics[width=2.5in,height=4.8in]
{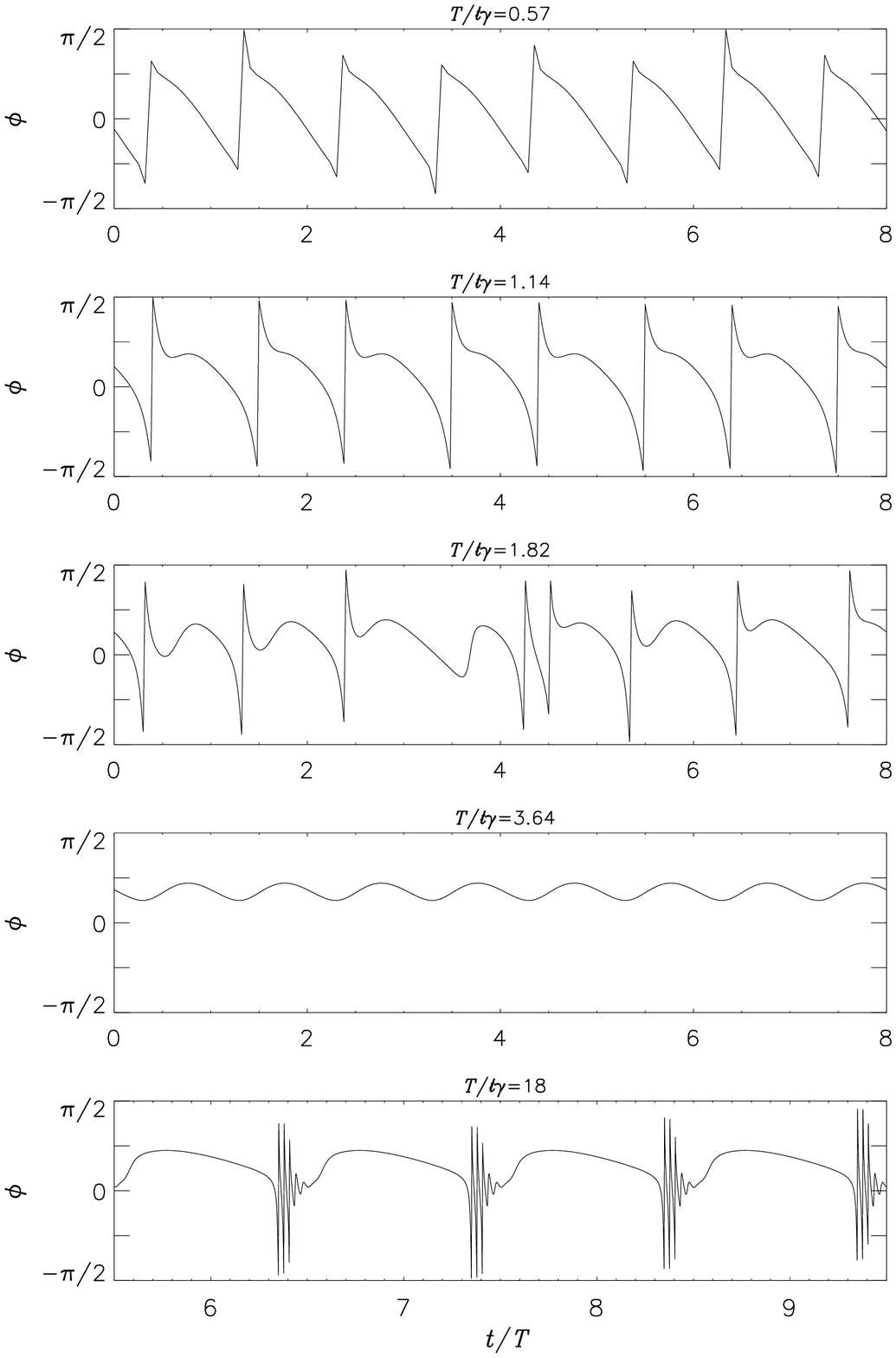} }
\end{array}
$
\caption{Same as figure \protect\ref{fig:chaos01}, except that the
results are from boundary-integral simulations and for slightly
different values of oscillation period.
Same $\betaMean$ and $\deltaBeta$ as in figure \ref{fig:chaos01}.
%
%
}
\label{fig:chaos03}
\end{figure}

A further increase of the period of the external forcing results in a
qualitative change of drop response.  We find that the drop still
undergoes a tumbling motion; however, the evolution is not periodic
but it becomes {\it chaotic}, as shown in figure
\ref{fig:chaos01}\subfig{c}.  The chaotic motion continues up to
$T/\tauDeformation=1.14$, 
and then the drop reverts to periodic motion. For $T/\tauDeformation$
(figure \ref{fig:chaos01}\subfign{d}) the drop oscillates about the
elongated stationary shape, and in the regime $T/\tauDeformation\gg1$
(figure \ref{fig:chaos01}\subfign{e}) the system approaches the
quasistatic behavior discussed in \S\,\ref{Hysteretic
behavior}.\footnote{The only significant deviation from the
quasistatic evolution for $T/\tauDeformation\gg1$ occurs right after
the drop jumps from the long to the compact shape when $\beta$
increases above the upper critical value $\betaUp$.  Namely, before
the drop settles down to the compact stationary shape it undergoes a
tumbling motion with the amplitude decaying on the timescale
\refeq{relaxation time scale}.} 

The transition to the chaotic drop motion occurs through a cascade of
period doubling events, as illustrated in figures \ref{example of
period doubling} and \ref{fig:period_doubling}.  Figure \ref{example
of period doubling} depicts the drop evolution at three values of the
amplitude of vorticity oscillations $\deltaBeta$ (the remaining system
parameters are the same as those that yield the chaotic motion
depicted in figure \ref{fig:chaos01}\subfign{c}).  The results shown
in figure \ref{example of period doubling}\subfig{a} indicate that for
a sufficiently small oscillation amplitude ($\deltaBeta=0.06$ in our
example) the drop evolves with the period $T_\drop=T$ (i.e., the
period equal to that of the external forcing).  At a larger amplitude
$\deltaBeta=0.08$ the drop oscillation period is $T_\drop=2T$ and for
$\deltaBeta=0.083$ we find $T_\drop=4T$.  The period-doubling scenario
of the transition to chaos in our system is further supported by the
bifurcation diagram 
shown in figure \ref{fig:period_doubling}, where the drop
length $\dropLength$ at times $t=nT$ ($n=1,2,\ldots$) is plotted
versus the flow-oscillation amplitude $\deltaBeta$.

An analysis of the results shown in figure \ref{example of period
doubling} indicates that the period doubling occurs as a result of a
resonance between the drop tumbling motion and the vorticity
oscillations.  Namely, if the drop is relatively long and
approximately aligned with the straining axis of the external flow
when the vorticity parameter $\beta(t)$ reaches a minimum, the drop
rotation may be significantly slowed down or even arrested (as in the
long-drop stationary state discussed in \S\,\ref{Hysteretic behavior}).
Such a temporary arrest of drop rotation corresponds to the shoulders
in the plots of the angular evolution depicted in figure \ref{example
of period doubling}.  On the other hand, if the drop angle exceeds
$\dropAngle=\pi/2$ when the vorticity goes through a minimum, the
arrest of drop rotation does not occur.  As seen in figure
\ref{example of period doubling}\subfig{b} this interplay of drop
tumbling with oscillations of the external forcing produces the
period-doubling bifurcation that leads to alternating accelerated and
retarded drop-rotation cycles.

In our numerical calculations depicted in figures
\ref{fig:chaos01}-\ref{fig:period_doubling} we have used the full set
of the small-deformation equations \refeq{small-deformation equations}
but we find that the simplified asymptotic equations \refeq{simplified
evolution equation} yield similar results.  In particular these
equations correctly reproduce the cascade of the period doubling
bifurcations and the chaotic-evolution domain.  However, a further
simplification of the evolution equations is not possible: if any of
the terms in equations \refeq{simplified evolution equation} is
removed the solutions qualitatively change and the chaotic domain
disappears.  We also find that chaotic drop dynamics occurs only for
highly viscous drops with $\lambda>200$.

\begin{figure}
\begin{center}
{\includegraphics[width=3.5in,height=3.in]{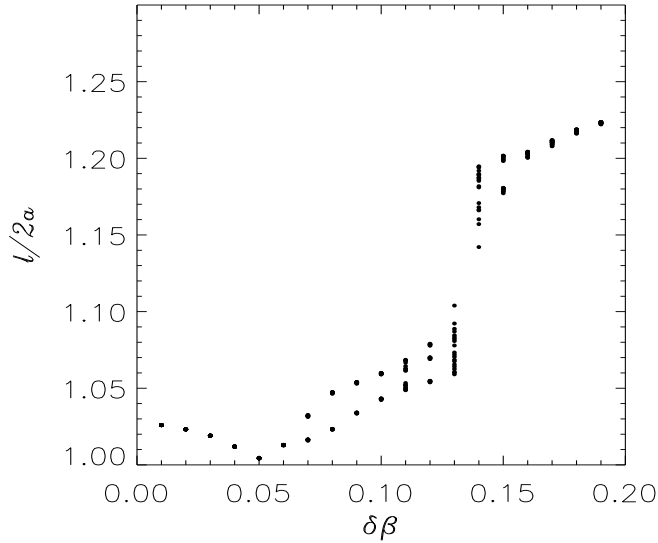}}
\end{center}
\caption{Bifurcation diagram for viscous drop in linear flow with
harmonic vorticity variation with mean $\bar{\beta}=0.22$,
$T/\tauDeformation=1.82$, $\lambda=275$, and $\Ca=0.2$. (Results from
boundary-integral simulations.)
%
%
}
\label{fig:period_doubling_dns}
\end{figure}

\subsection{Boundary-integral results}
\label{subsec:chaos_bis}

Figure \ref{fig:chaos03} shows examples of our simulation results for
the periodic and chaotic drop evolution.  The flow parameters are
similar to those used in our small-deformation calculations described
in \S\,\ref{subsec:chaos_sdt}.  As with the small-deformation
calculations, for short periods of the external forcing $T$ the drop
oscillates around the compact stationary shape, for moderate periods
the system undergoes a transition to chaotic evolution, and for long
periods the drop motion approaches the quasistatic behavior.
Consistent with the small-deformation results, the chaotic drop
dynamics revealed by the boundary-integral simulations is due to a
cascade of period doubling bifurcations.  A bifurcation diagram illustrating
this behavior is presented in figure \ref{fig:period_doubling_dns}.

The domain of chaotic motion found in the direct boundary-integral
simulations somewhat differs from the corresponding domain obtained
from the small-deformation theory.  Also, the magnitude of the chaotic
fluctuations in the drop length is larger in the boundary-integral
runs.  We have tested the convergence of the boundary-integral
simulations, and we believe that the differences between the drop
behavior obtained by the two different methods stem from the
approximations involved in the small-deformation theory.  We note,
however, that the evolution in the chaotic and period-doubling regimes
is very sensitive to the to small changes of the system parameters
(hence, also sensitive to the approximations involved in our
calculations).

\begin{figure}
\begin{center}
{\includegraphics[width=3.5in,height=3.5in]{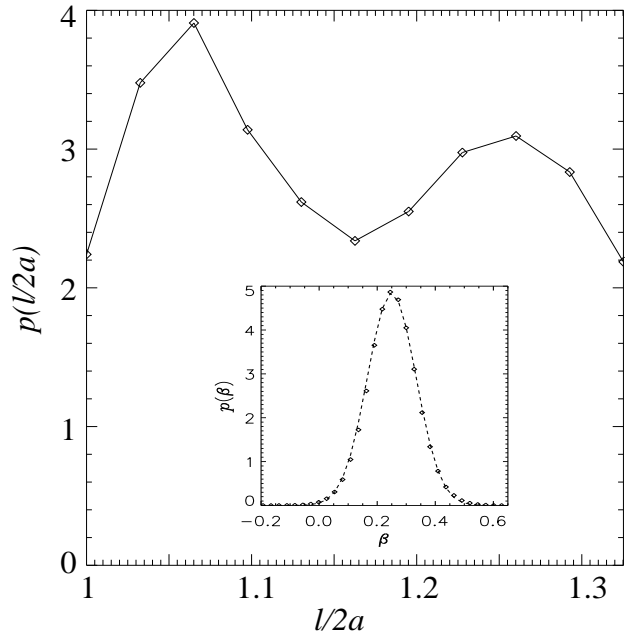}}
\end{center}
\caption{
Length probability distribution for viscous drop with $\lambda=200$
and $\Ca=0.2$, in linear flow with stochastic vorticity.  Mean value
of the vorticity parameter $\betaAver=0.25$, variance
$\betaVariance=0.13$, and correlation time
$\tauCorrelation/\tauDeformation=0.24$. Inset shows
vorticity probability distribution.  (Results from boundary-integral
simulations.)}
\label{fig:CGaussdns}
\end{figure}
\begin{figure}
{\includegraphics[width=2.6in,height=2.2in]{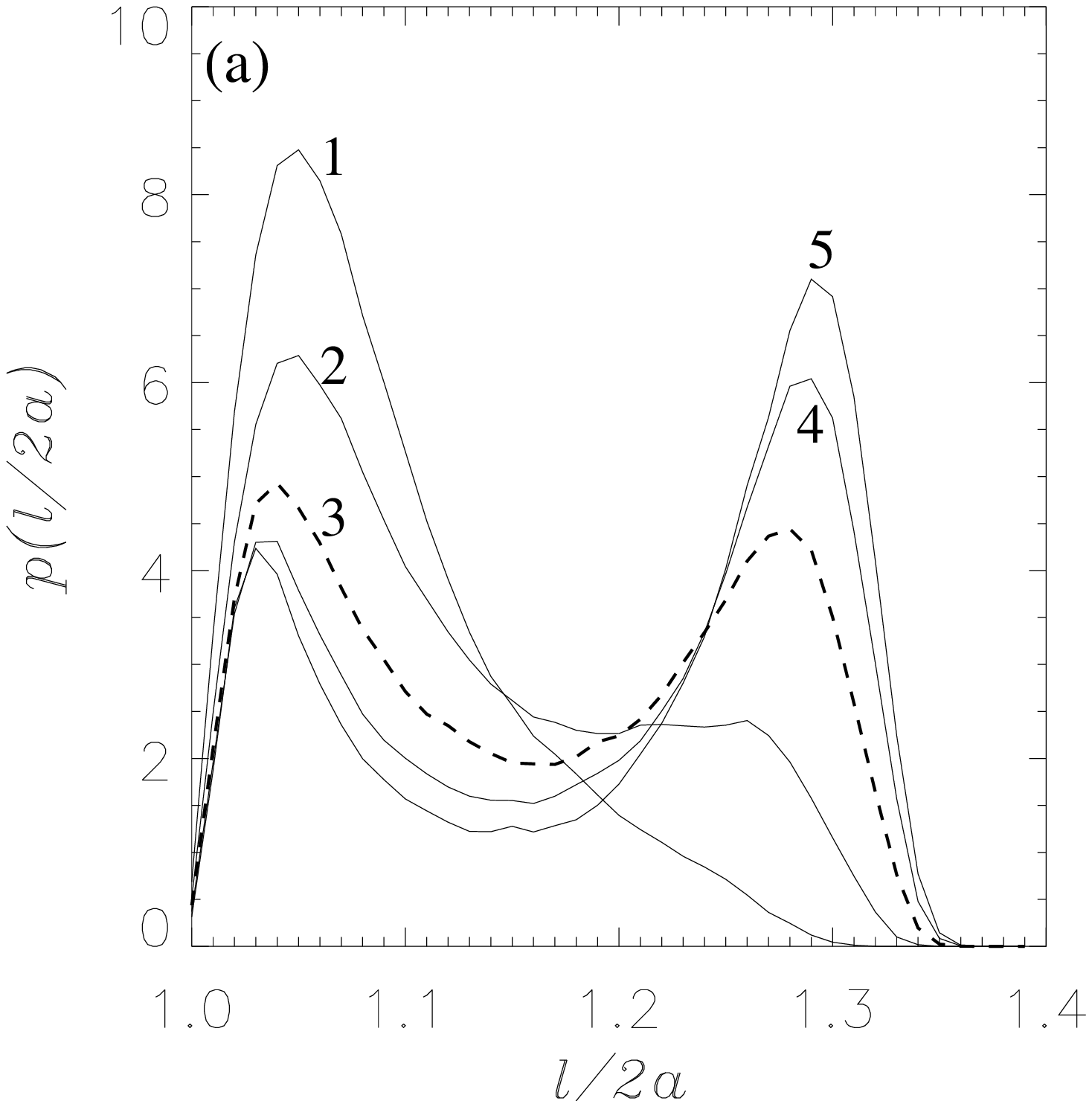}}
{\includegraphics[width=2.6in,height=2.2in]{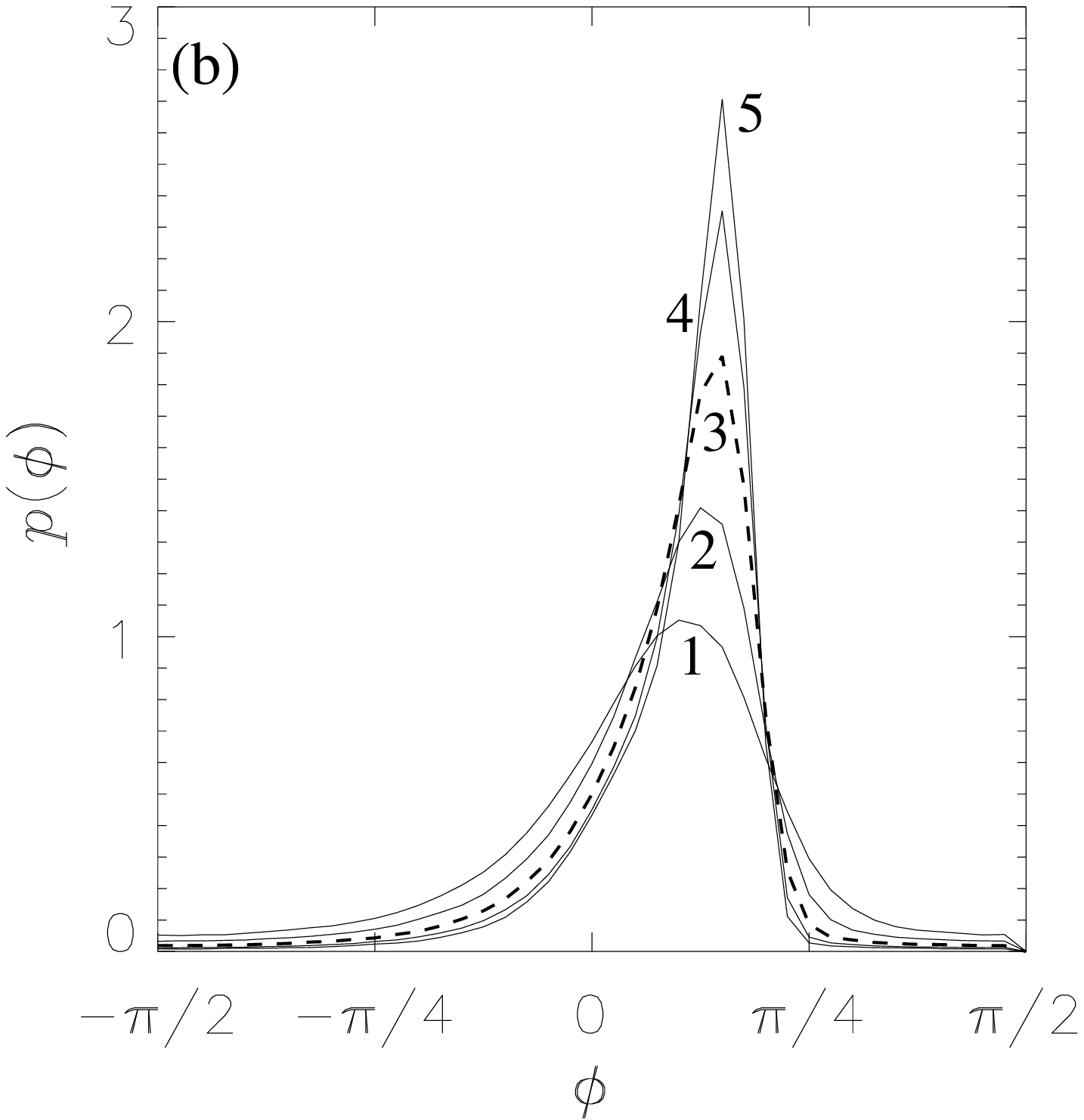}}
\caption{
\subfig{a} Length and \subfig{b} angle probability distributions for
viscous drop with $\lambda=200$ and $\Ca=0.2$, in linear flow with
stochastic vorticity.  Mean and variance of the vorticity parameter
are the smae as in figure \protect\ref{fig:CGaussdns}; the correlation
times are $\tauCorrelation/\tauDeformation=0.025,0.1,0.2,0.3,0.4$
for lines marked 1--5, respectively.
(Results from small-deformation theory.)
%
%
}
\label{fig:CGauss01}
\end{figure}

\section{Drop Statistics in a linear flow with Stochastic Vorticity}
\label{Drop Statistics in a linear flow with Stochastic Vorticity}

In some systems (e.g. emulsion flows through a packed bed of fibers
\citep{Mosler-Shaqfeh:1997} or turbulent emulsion flows with drops
that are much smaller then the Kolmogorov scale
\citep{Cristini-Blawzdziewicz-Loewenberg-Collins:2003}) a viscous drop
undergoes deformation in a random external creeping flow.  To gain
some understanding of the role that drop bistability may play in such
systems we consider the drop behavior in a flow with stochastic
vorticity.

We assume that the time variation of the vorticity parameter $\beta$
is described by a stationary Markovian Gaussian process (i.e, the
Ornstein--Uhlenbeck process) with the mean $\betaAver$, variance
$\betaVariance$, and correlation time $\tauCorrelation$.  A standard
numerical scheme \citep{Fox-Gatland-Roy-Vemuri:1988} for generating
such a time-correlated Gaussian process is used to model the time
variation of $\beta$ for a given drop trajectory.

An example of drop behavior in a stochastic flow with a Gaussian
variation of the vorticity is presented in figure \ref{fig:CGaussdns}.
The mean value of the vorticity is $\betaAver=0.25$ and the variance
is $\betaVariance=0.13$.  The correlation time of the vorticity
distribution $\tauCorrelation/\tauRelaxation=0.24$ is several times
shorter than the drop relaxation time.  The figure depicts the
probability distribution of the drop length, obtained using the
boundary-integral simulations.  The capillary number is $\Ca=0.2$,
and the drop viscosity is $\lambda=200$.  

The results indicate that the drop-length distribution is bimodal for
the above parameter values.  This behavior is expected since the
vorticity undergoes random variation in the domain that includes both
the lower and upper critical values $\betaLow$, and $\betaUp$ of the
vorticity parameter $\beta$.  Since drop response to a flow with slow
variation of vorticity is hysteretic, a drop in the random flow tends
to stay in the neighborhood of the compact and the elongated
stationary states.

We note that the peak of the length probability distribution at
$\dropLength\approx1.25$ is shifted towards the shorter drop lengths
compared to the length of a drop in the elongated stationary shape.
This is because $\Ca=0.2$ is close to the critical capillary number
for drop breakup.  Thus, due to the slow time scale in the drop
dynamics near the critical capillary number
\citep{Blawzdziewicz-Cristini-Loewenberg:2002a}, the drop does not
have sufficient time to fully extend before the vorticity
significantly changes.

Figure \ref{fig:CGauss01} shows the probability density distribution
for the drop length $\dropLength$ and drop angle $\dropAngle$ for
different values of the flow correlation time $\tauCorrelation$.
Other system parameters are the same as in figure \ref{fig:CGaussdns}.
The calculations were performed using the small-deformation equations
\refeq{small-deformation equations}.  The results indicate that at
short flow correlation times the drop-length probability distribution
is peaked around small values corresponding to the short-drop
stationary solution, and the has a moderate-hight peak at
$\phi\approx\pi/4$.  As the flow correlation time $\tauCorrelation$
increases, the length probability distribution becomes bimodal: one of
its peaks corresponds to compact and the other to elongated drops.  A
corresponding change occurs in the angle distribution, i.e., its peak
becomes more pronounced, and shifts towards the straining axis
$\phi=\pi/4$.

\begin{figure}
{\includegraphics[width=2.6in,height=2.2in]{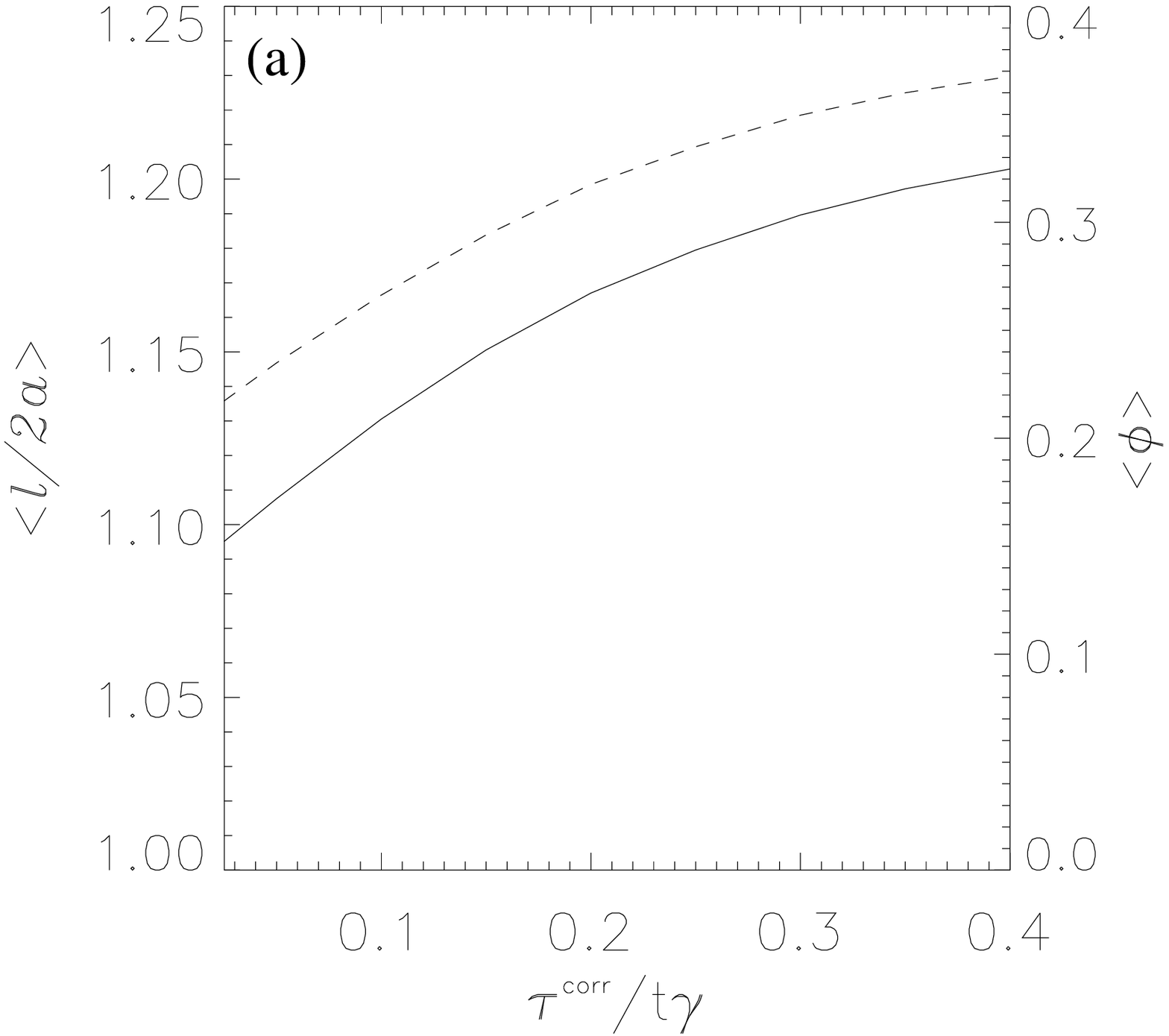}}
{\includegraphics[width=2.6in,height=2.2in]{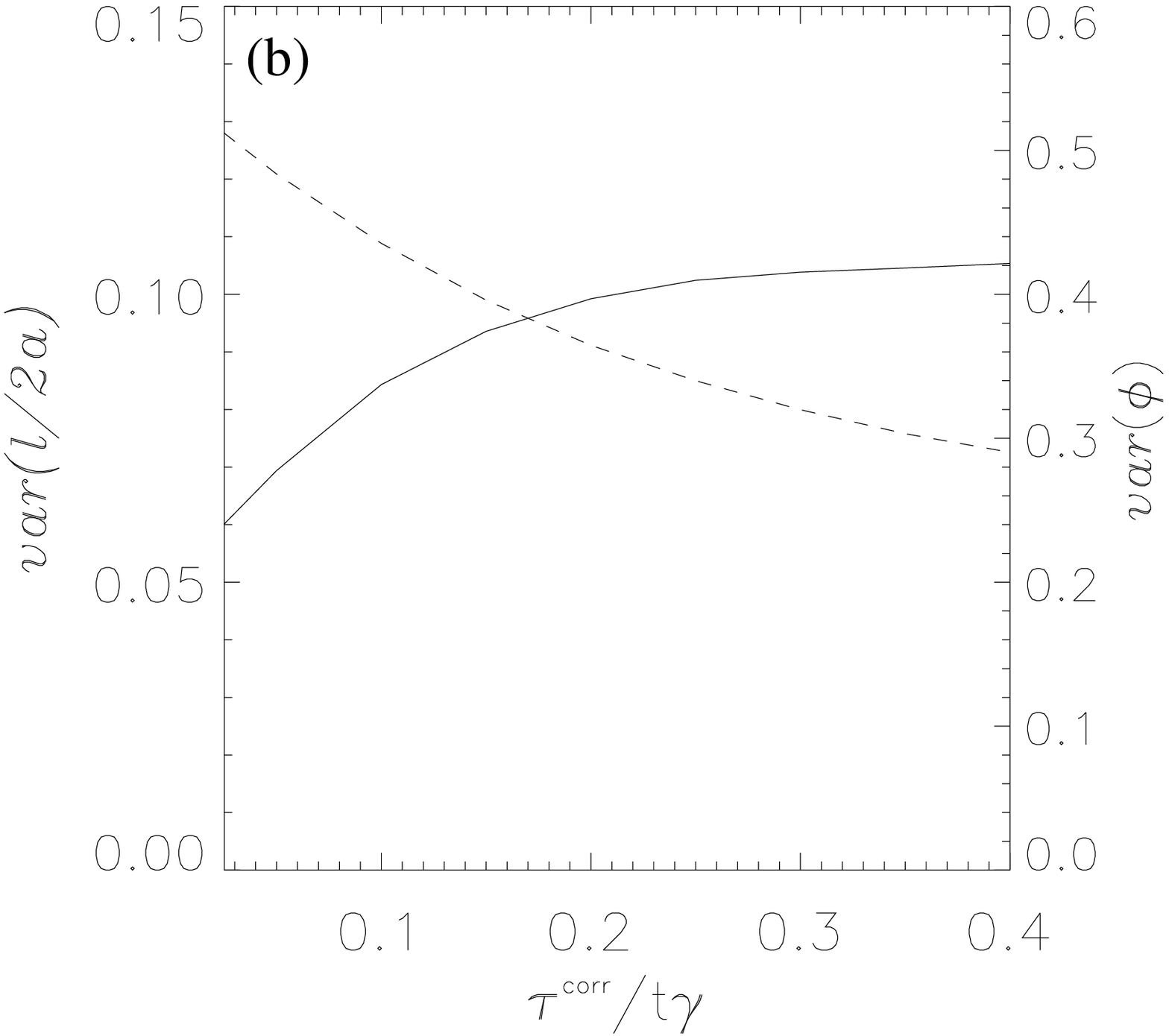}}
\caption{
\subfig{a} Mean and \subfig{b} variance of drop length (dashed) and
angle (broken line) versus correlation time $\tauCorrelation$
normalized by drop deformation time, for $\lambda=200$, $\Ca=0.2$,
$\betaAver=0.25$, and $\betaVariance=0.13$.
}
\label{fig:CGauss02}
\end{figure}

The shift of the typical drop length and orientation from the compact
to elongated state when the flow correlation time is increased
resembles the analogous shift for a system with harmonic vorticity
oscillations (see figures \ref{fig:chaos01}\subfign{b} and figures
\ref{fig:chaos01}\subfign{d}).  This behavior is further illustrated
in figure \ref{fig:CGauss02} which shows the average values and the
variance of $\dropLength$ and $\dropAngle$ versus the correlation time
$\tauCorrelation$.

\section{Conclusions}
\label{sec:conclusion}

We have presented results of numerical and theoretical investigations
of the dynamics of highly viscous drops in two-dimensional linear
creeping flows with time-dependent vorticity.  In our earlier
publication \citep{Blawzdziewicz-Cristini-Loewenberg:2003} we
predicted that in stationary flows such drops exhibit bistable
behavior: there is a range of system parameters where the drop may
assume either an elongated shape approximately aligned with the
straining axis of the external flow or a nearly spherical shape
approximately aligned in the flow direction.  Here we analyze the
consequences of this behavior for the system dynamics.  We also
elucidate the physical mechanism that leads to drop bistability.

A direct consequence of the existence of two stationary states is
hysteretic drop response to a flow with slowly varying vorticity.  We
have explained that the rapid transition from an elongated
non-rotating drop shape to the nearly spherical compact shape occurs
when the vorticity becomes strong enough to overcome the effect of the
straining flow component that aligns the drop with the straining axis.
This transition is thus analogous to the behavior of an elongated
rigid particle which starts to tumble when the vorticity grows above a
critical value \citep{Jeffery:1922}.  A viscous drop also begins to
tumble at a critical vorticity magnitude $\betaUp$.  However, when the
drop becomes misaligned with the extensional axis it relaxes under the
action of capillary forces towards a nearly spherical rotationally
stabilized stationary shape.

If, in turn, the vorticity is slowly decreased, the drop returns to
the elongated shape only after the vorticity magnitude reaches a lower
critical value $\betaLow<\betaUp$.  This hysteretic drop response
occurs at high drop viscosities because the rotational stabilizing
mechanism is more efficient in the high-viscosity regime.  A highly
viscous drop deforms less within each drop revolution, so the compact
shape remains stable even for small vorticity magnitudes.

The existence of two stationary states affects drop dynamics not only
in the quasistatic regime but also at finite frequencies of the
external forcing.  At small amplitudes of harmonic vorticity
oscillations the drop simply oscillates (with the same frequency as
the external forcing) about one of the stationary states.  However, if
the vorticity-variation range includes both critical values $\betaLow$
and $\betaUp$ the dynamics of the system is much richer.  We find that
with an increasing magnitude of the vorticity oscillations the system
undergoes a cascade of period-doubling bifurcations resulting in
chaotic drop dynamics.  The period doubling stems from the resonance
between the periodicity of the external forcing and the tumbling
motion of the drop when it jumps from a (partially) elongated shape
towards the compact rotationally stabilized state.

Chaos in our system emerges despite linearity of Stokes equations --
the system dynamics is nonlinear because of the coupling of the flow
to the evolving fluid interface.  A detailed analysis of
small-deformation equations describing drop dynamics reveals that in
addition to chaos associated with the period-doubling mechanism there
also exists in our system a different kind of chaotic evolution that
results from manifold tangling \citep{Guckenheimer-Holmes:1983}.  Our
analysis of different types of chaos  will be presented in a separate
publication.

To our knowledge, chaotic drop dynamics in Stokes-flow regime has
never been observed before.  We note, however, that in a recent
independent study \citep{Papageorgiou:2007} has reported chaotic
dynamics in co-annular Stokes flow with insoluble surfactant adsorbed
on the fluid interface. Chaos in their system also appears as a result
of period doubling.

Understanding of drop bistability and the associated dynamical
phenomena is relevant for many practical problems.  For example,
interpretation of rheological response of emulsions of highly viscous
drops to time-varying flows requires insight into drop dynamics.  Our
results may also be useful in design new methods for manipulating
emulsion microstructure in material processing and controlling drop
behavior in microfluidic flows.  Drop bistability could, e.g., be used
to construct microfluidic switches, and chaotic drop dynamics may be
relevant for microfluidic mixing.

The stabilizing and destabilizing mechanisms described in our paper
apply not only to viscous drops but also to other deformable
particles.  Therefore, results of our study have a broader
significance.

In particular, our analysis suggests that macromolecules with high
degree of internal dissipation may undergo a transition between a
nearly spherical and moderately elongated states \cite[in addition to
the standard coil-stretch transition predicted by][]{deGennes:1974}.
In fact, the dynamics of macromolecules can be modeled using equations
analogous to \refeq{simplified evolution equation}, supplemented with
terms representing random thermal forces.  Such a simplified
description correctly captures the most important features of power
spectra of DNA molecules evolving in linear flows with nonzero
rotational component \citep{Blawzdziewicz:2006}.

There are also close analogies between drop and vesicle motion. The
main difference between these two systems is that vesicles satisfy a
constant-area constraint whereas the drop area can vary.  This
constraint gives rise to periodic vesicle motion (such as tank
treading and tumbling) even in stationary flows
\citep{Misbah:2006,Vlahovska-Gracia:2007}.  It would be interesting to
determine if a coupling of vesicle oscillations to a harmonic
variation of the external flow can lead to chaotic dynamics.

It would also be of significant interest to experimentally explore the
bistable and chaotic drop dynamics (as well as related phenomena for
other deformable particles).  In such experiments a four-roll mill
could be used to produce a linear flow with a controlled magnitude of
vorticity.  The experiments could also be performed using recently
developed microfluidic analogues of a four-roll mill device
\citep{%
Hudson-Phelan-Handler-Cabral-Migler-Amis:2004,%
Lee-Dylla_Spears-Teclemariam-Muller:2007%
}.

\begin{acknowledgments}
We would like to acknowledge helpful discussions with Paul Steen,
Demetrios Papageorgiou, and Michael Loewenberg.  We thank Michael
Loewenberg for permission to use his improved boundary-integral code,
and also Petia Vlahovska for use of her Mathematica codes for the 
coefficients in the small-deformation theory.
JB was supported by NSF CAREER grant CTS-0348175, YNY acknowledges a
NSF/DMS grant (DMS-0708977) and a SBR grant from NJIT.  The simulations were
conducted on the NJIT computer cluster supported by NSF/MRI grant
DMS-0420590.
\end{acknowledgments}

\end{document}